\documentclass[12pt,letterpaper]{article}
\usepackage{amssymb}
\usepackage{amsmath}
\usepackage{amsfonts}
\usepackage{epsfig}
\usepackage{feynmf}
\usepackage{color}
\usepackage{hyperref}

\newcommand{\sect}[1]{\setcounter{equation}{0}\section{#1}}

\usepackage{bm}
\outer\def\beginsection#1\par{\medbreak\bigskip
      \message{#1}\leftline{\bf#1}\nobreak\medskip
\vskip-\parskip
      \noindent}

\newcommand{\eq}{\begin{equation}}
\newcommand{\eqx}{\end{equation}}
\newcommand{\eqn}{\begin{eqnarray}}
\newcommand{\eqnx}{\end{eqnarray}}
\newcommand{\bi}{\begin{itemize}}
\newcommand{\ei}{\end{itemize}}

\def\Ord{{\cal O}}

\def\q{q}
\def\be{\begin{equation}}
\def\ee{\end{equation}}
\def\ba{\begin{eqnarray}}

\newcommand{\ea}[1]{\begin{align} #1 \end{align}}

\newcommand{\aqk}[1]{ \alpha' q k_{#1}}

\newcommand{\zz}[2]{z_{#1}-z_{#2}}

\usepackage[titles]{tocloft}

\begin{document}
\begin{titlepage}
\hfill \hbox{NORDITA-2015-021}
\vskip 1.5cm
\vskip 1.0cm
\begin{center}
{\Large \bf  Soft theorem for the graviton, dilaton and the Kalb-Ramond field in the bosonic string }
 
\vskip 1.0cm {\large Paolo
Di Vecchia$^{a,b}$,
Raffaele Marotta$^{c}$, Matin Mojaza$^{b}$ } \\[0.7cm] 
{\it $^a$ The Niels Bohr Institute, University of Copenhagen, Blegdamsvej 17, \\
DK-2100 Copenhagen, Denmark}\\
{\it $^b$ Nordita, KTH Royal Institute of Technology and Stockholm University, \\Roslagstullsbacken 23, SE-10691 Stockholm, Sweden}\\
{\it $^c$  Instituto Nazionale di Fisica Nucleare, Sezione di Napoli, Complesso \\
 Universitario di Monte S. Angelo ed. 6, via Cintia, 80126, Napoli, Italy}
\end{center}
\begin{abstract}
We study the  behavior of the scattering amplitudes of the bosonic string involving a  soft massless state (graviton, dilaton and Kalb-Ramond antisymmetric tensor)  and closed string tachyons or other closed string massless states.  For a soft graviton we confirm the results, obtained in 
Ref.~\cite{BDDN}  using just gauge invariance, up to terms of $\Ord (q^1)$ for external tachyons and up to terms of $\Ord (q^0)$ for external massless closed string states. Furthermore, we also derive the behavior of the scattering amplitude when a dilaton or a Kalb-Ramond field becomes soft.
These results are new and 
cannot, to our knowledge, be derived by using gauge invariance. 
It turns out, in the cases examined, that the soft  amplitude for a dilaton  or   for an  antisymmetric tensor is obtained  by saturating the tensor, $M_{\mu \nu}$, derived from gauge invariance for gravitons, with their respective polarization tensors.
Thus extra terms that could have appeared in $M_{\mu \nu}$ in the case of a soft dilaton, in fact do not appear.

\end{abstract}
\end{titlepage}

\tableofcontents

\section{Introduction and results}
\label{intro}

Soft theorems have a long history and were studied already in the 1950s and 1960s, especially for Compton scattering of photons and gravitons on arbitrary targets (for a discussion of low-energy theorem for photons see Chapter 3 of Ref. \cite{DFFR}). They were recognized to be important consequences of local gauge invariance~\cite{LowFourPt, LowTheorem, Weinberg, OtherSoftPhotons}.  For photons, Low's
theorem~\cite{LowTheorem} determines the amplitudes with a soft
photon from the corresponding amplitudes without a photon, up  to terms of order
$\Ord(q^0)$, where $q$ is the soft-photon momentum.  
The universal  leading behavior of a soft-graviton was first discussed by Weinberg~\cite{Weinberg}. 
Non-leading terms were then discussed in Refs.~\cite{GrossJackiw,Jackiw}.  More recent discussions  of
the generic subleading behavior of soft gluons and gravitons are given
in Refs.~\cite{WhiteYM,WhiteGrav}.

In the 1970s soft theorems for the string dilaton were discussed by Ademollo et al.~\cite{Ademollo:1975pf} and by Shapiro~\cite{Shapiro:1975cz}
for tree diagram scattering amplitudes involving only massless particles~\footnote{See also Refs.~\cite{Yoneya:1987gc,Hata:1992it} for the study of the soft dilaton behavior in string field theory.}. 
Gauge invariance does not in general determine the soft behavior of the string dilaton, because of the potential presence of gauge invariant terms at order $\Ord(q^0)$. It was nevertheless found in Ref.~\cite{Ademollo:1975pf} that 
such terms are not present in the bosonic string, if the amplitude involves only massless closed string states.
They do, however, appear if massless open string states are also involved.

Interest in the soft behavior of gravitons and gluons has recently been 
renewed by a proposal from Strominger and
collaborators~\cite{Strominger,CachazoStrominger} showing that
the soft-graviton behavior follows from Ward identities of extended Bondi,
van der Burg, Metzner and Sachs (BMS) symmetry~\cite{BMS,
  ExtendedBMS}.  
   This has stimulated the study of the subleading soft 
 behavior in amplitudes with gluons and gravitons. In four spacetime
dimensions, Cachazo and Strominger~\cite{CachazoStrominger} proposed 
that tree-level graviton amplitudes have a universal behavior through
second subleading order in the soft-graviton momentum. These  considerations have since been extended to gluons in arbitrary number of  
dimensions in various ways Ê at tree level~\cite{SoftGluonProof,Volovich,Conformal,IntegrandSoft,TwistorSoft}.

Poincar{\'{e}} and gauge invariance as well as a condition arising from the distributional nature of scattering amplitudes have been used  in Ref.~\cite{NewPaper} to  strongly constrain the soft behavior for gluons and gravitons, while in Ref.~\cite{BDDN} gauge invariance is shown to completely fix the first two   
leading terms  (up to terms $\Ord (q^0)$) in the case of a gluon, and the first three leading terms (up to terms $\Ord (q^1)$) in the case of a graviton, for any number 
 of space-time dimensions ($q$ being the soft momentum). 
 More specifically, in Ref.~\cite{BDDN} it was shown that by imposing the conditions
\begin{eqnarray}
q_{\mu} M^{\mu \nu}_{n+1} =  q_{\nu} M^{\mu \nu}_{n+1} =0 \ ,
\label{qqM}
\end{eqnarray}
one could determine for small $q$ the behavior of the on-shell amplitude  $M^{\mu \nu}_{n+1}$ 
containing a graviton and $n$ scalar particles
\footnote{$\kappa_d$ is connected to Newton's gravitational constant  $G_N^{(d)}$ by the relation $ \kappa_d \equiv \sqrt{8\pi G_N^{(d)}}$, where $d$ is number of space-time dimensions.}: 
\begin{eqnarray}
M^{\mu \nu}_{n+1} &=&  \kappa_d\sum_{i=1}^{n} \frac{1}{ k_i\q}\left[  k_i^{\mu} k_i^{\nu}   - \frac{i}{2}  k_i^{\nu} q_{\rho} J_i^{\mu \rho}   - \frac{i}{2}  k_i^{\mu} q_{\rho} J_i^{\nu \rho}  - \frac{i}{2} q_{\rho} q_{\sigma} \left(
k_i^{\nu} J_i^{\mu \rho} \frac{\partial}{\partial k_{i\sigma}} - k_i^{\sigma} J_i^{\mu \rho} \frac{\partial}{\partial k_{i\nu}}   \right) \right]  T_n
\nonumber \\
&=& \kappa_d \sum_{i=1}^{n} \frac{1}{ k_i\q}\left[  k_i^{\nu} k_i^{\mu}   - \frac{i}{2} k_i^{\nu} q_{\rho} J_i^{\mu \rho}    - \frac{i}{2} k_i^{\mu} q_{\rho} J_i^{\nu \rho}  - \frac{1}{2} q_{\rho}J_i^{\mu \rho} q_{\sigma}  J_i^{\nu \sigma} \right. \nonumber \\
 && 
 \qquad \qquad  \quad
 +  \left. \frac{1}{2} \Big((k_i q)( \eta^{\mu \nu} q^{\sigma} -  q^{\mu} \eta^{\nu\sigma} ) - k_i^{\mu} q^{\nu} q^{\sigma}\Big) \frac{\partial}{\partial k_i^{\sigma}} \right] T_n
  + \Ord (q^2) \ ,
\label{MMMmunu}
\end{eqnarray}
where $T_n$ is the amplitude of $n$ scalar particles  and $J_i^{\mu \rho}$ is the angular momentum operator
\begin{eqnarray}
J_i^{\mu \rho} = i \left( k_{i}^{\mu} \frac{\partial}{\partial k_{i\rho} }-  k_{i}^{\rho} \frac{\partial}{\partial k_{i\mu}}   \right) \, .
\label{Jmurho}
\end{eqnarray}
Actually, as we will  see later,  the conditions in Eq.~(\ref{qqM}) are in general not correct, because the right hand side  can contain  a  term proportional to the momentum $q$. However, in the case of a graviton, this extra term  is  irrelevant.

Before proceeding further note  that  next-to-leading soft-graviton theorems in arbitrary number of  
dimensions were also studied in Refs.~\cite{Royas,Zlotnikov,LNS,CHW,SabioVera}
and that soft gluon and graviton behaviors are in general modified by loop corrections, as discussed in Refs.~\cite{LoopCorrections,HeHuang,BDDN,DDLPR}. Finally, soft gluon and graviton behavior were also studied in the framework of superstring theory in Refs.~\cite{IntegrandSoft,StringSoft,Schwab}.  In particular, in the previous  papers it has been shown that the string amplitudes reproduce the  soft graviton  behavior discussed above up to the first subleading term in the soft momentum without any extra $\alpha'$ corrections.

In this paper we concentrate on the closed bosonic string and we study the soft behavior of a scattering amplitude involving one soft massless state with the other states being either closed string tachyons or other massless closed string states.  The aim of this paper is not only to check that in the case of a soft graviton, one obtains a  soft behavior  consistent  with what is required by gauge invariance, as discussed in Ref.~\cite{BDDN}, but also and especially to get 
the soft behavior for the  dilaton and for the  Kalb-Ramond field, which is  not obvious how to obtain in field theory.  

The low-energy field theory action that  in the string frame describes  the interaction
between gravity, dilatons and Kalb-Ramond fields in $d$ dimensions reads 
\footnote{In this paper we keep the number of space-time dimensions $d$ arbitrary, but for the bosonic string it is implied that $d=26$.}
\begin{eqnarray}
S_{\rm string} = \frac{1}{2 {\hat{\kappa}}_d^2} \int d^d x \sqrt{- G } \,\,\, {\rm e}^{-2 \phi} \left[ R + 4 G^{\mu \nu} \partial_{\mu} \phi \partial_{\nu} \phi - \frac{1}{2 \cdot 3!}   H_{\mu \nu \rho}^2 \right] \ .
\label{Sstring}
\end{eqnarray}
The corresponding action in the Einstein frame becomes~\footnote{The relation between the metric in the string and in the Einstein frame is given by $g_{\mu \nu} = {\rm e}^{- \frac{4}{d-2} \phi } G_{\mu \nu}$.} :
\begin{eqnarray}
S_{\rm E} = \frac{1}{2 \kappa_d^2} \int d^d x \sqrt{- g} \left[ R - \frac{1}{2} g^{\mu \nu} \partial_{\mu} {\hat{\phi}} \partial_{\nu} {\hat{\phi}}  - \frac{1}{2 \cdot 3!} {\rm e}^{-\sqrt{\frac{8}{d-2}} {\hat{\phi}} }  \left (H_{\mu \nu \rho} \right )^2 \right] \ ,
\label{Seffect}
\end{eqnarray}
where ${\hat{\phi}}$ is the dilaton field (canonically normalized)\footnote{$\phi$ is related to ${\hat{\phi}}$ by the relation ${\hat{\phi}} = \sqrt{\frac{8}{d-2}} \phi$, while $\kappa_d^2 = {\hat{\kappa}}_d^2 {\rm e}^{2 <\phi>}$.}  and
\begin{eqnarray}
H_{\mu \nu \rho} = \partial_{\mu} B_{\nu \rho} - \partial_{\nu} B_{\mu \rho} + \partial_{\rho} B_{\mu \nu} \ ,
\label{Hmunurho}
\end{eqnarray}
which is antisymmetric under the exchange of the three indices since  $B_{\mu \nu}$, the Kalb-Ramond field, is also antisymmetric.  From these actions it is not obvious at all that there is a gauge invariance determining the soft behavior of amplitudes with a soft dilaton, similar to the one for gravitons. 
It is also not clear how to get a low-energy theorem for the Kalb-Ramond field by using its gauge invariance; i.e. in the case of the graviton amplitudes,  the subleading behaviour of the amplitudes in the soft momentum expansion, is related to the leading one by gauge invariance.  For the Kalb-Ramond field, however, the leading term is absent and this procedure apparently seems to fail.

The reason why  this is instead possible in string theory, is due to the fact that the scattering amplitudes involving a graviton or a dilaton or a Kalb-Ramond field with momentum $q$  and other particles with momentum $k_i$, are all obtained  from the same two-index  tensor $M^{\mu \nu} (q; k_i)$ by saturating it with   a polarization tensor satisfying
respectively the following conditions:
\begin{subequations}
\label{GdB}
\begin{eqnarray}
\text{Graviton}\, \, (g_{\mu \nu})  \,\,\, &\Longrightarrow& \epsilon^{\mu \nu}_g = \epsilon^{\nu \mu}_g  \,\,\, ; \,\,\, \eta_{\mu \nu} \epsilon^{\mu \nu}_g =0  \label{epsG} \\
 \text{Dilaton } \, (\phi)  \,\,\, &\Longrightarrow&   \epsilon^{\mu \nu}_d =  \eta^{\mu \nu} - q^{\mu} {\bar{q}}^{\nu}  - q^{\nu} {\bar{q}}^{\mu} \label{epsd} \\
 \text{Kalb-Ramond }(B_{\mu \nu} )  \,\,\,  &\Longrightarrow&   \epsilon^{\mu \nu}_B = - \epsilon^{\nu \mu}_B 
\label{epsB}
\end{eqnarray}
\end{subequations}
where ${\bar{q}}$ is, similarly to $q$, a lightlike vector such that $q \cdot {\bar{q}}=1$.   This is also what one gets, at least in the field theory limit,  by applying at tree level the KLT relations~\cite{Kawai:1985xq} or in general  the BCJ rules~\cite{Bern:2008qj, Bern:2010ue} according to which one also obtains  a common $M^{\mu \nu}$ that contains the soft behavior for  the graviton, dilaton and Kalb-Ramond field.  This is another example, where the scattering amplitude tells us more than the original Lagrangian. 

Furthermore, the tensor $M^{\mu \nu} (q; k_i)$ satisfies in general the following conditions:
\begin{eqnarray}
q^{\mu} \left( M_{\mu \nu} (q; k_i) - f (q;k_i) \eta_{\mu \nu} \right)  =0  \ , \ \
 q^{\nu} \left(M_{\mu \nu} (q; k_i) -  f (q;k_i)  \eta_{\mu \nu} \right) =0 \ .
\label{qM}
\end{eqnarray}
In the case of a graviton and of a Kalb-Ramond field, the term with the metric tensor $\eta_{\mu \nu}$ is irrelevant and the previous conditions reduce to those in Eq.~(\ref{qqM}), which in the case of a graviton fix the three leading terms in the limit of small $q$, as given in Eq.~(\ref{MMMmunu})
when the other particles are scalars.  The extra   term in Eq.~(\ref{qM}) 
is, however, relevant in the case of the dilaton. Thus in general the soft limit for the dilaton cannot be obtained as in the case of the graviton. One can, however, explicitly compute the scattering amplitude involving one soft dilaton and see if the extra terms proportional to $\eta_{\mu \nu}$ are  appearing  or not.  From Ref.~\cite{Ademollo:1975pf} it is already known that such extra terms are not present, up to terms of $\Ord (q^0)$, if the other states are  
massless closed ones. On the other hand, they are present if there are massless open string states.

In this work, we first  study  the soft behavior of a massless closed string state in an amplitude involving an arbitrary number of closed string tachyons. By explicitly performing the soft limit, we show that  the amplitude for small $q$ behaves  as follows:
\ea{
 M_{\mu \nu} (q; k_i ) =    
\kappa_d  \sum_{i=1}^{n} \Bigg[ 
 &\frac{k_{i\mu} k_{i\nu}}{k_i q} - i \frac{k_{i\nu} q^\rho J^{(i)}_{\mu \rho} }{2 k_i q} -  i \frac{k_{i\mu} q^\rho J^{(i)}_{\nu \rho} }{2 k_i q}  - \frac{1}{2} \frac{q_{\rho} J_i^{\mu \rho} q_{\sigma} J^{\nu \sigma}_i  }{k_i q}   \nonumber \\
& + \frac{1}{2} \Big(  \left(\eta^{\mu \nu}  q^\sigma - q^\mu \eta^{\nu \sigma}\right) - \frac{k_{i}^{\mu} q^\nu q^\sigma  }{ k_iq} \Big) \frac{\partial}{\partial k_i^{\sigma}} \Bigg] T_n (k_i)  + O(q^2) \ ,
\label{1gra4tacbvv}
}
where $T_n$ is the amplitude with $n$ closed string tachyons:
\ea{
T_n (k_i )  = \frac{8\pi}{\alpha'} \left( \frac{\kappa_d}{2 \pi} \right)^{n-2}
 \int \frac{\prod_{i=1}^{n} d^2 z_i}{d V_{abc}} \prod_{i \neq j} |z_i - z_j|^{\frac{\alpha'}{2} k_i k_j} \ ,
\label{Mtachyons}
}
with  the factor in front providing  the correct normalization for the $n$-tachyon amplitude~\footnote{The overall normalization in Eq.~(\ref{Mtachyons}) corresponds to  choosing $d^2 z_i = 2d (Re z_i) d (Im z_i)$ and also $d^2 z = 2 d (Re z) d (Im z)$ .}.
Notice that Eq. (\ref{1gra4tacbvv}) has precisely the same form as Eq. (\ref{MMMmunu}) without any additional term proportional to $\eta^{\mu \nu}$ and $\alpha'$  corrections.

In the case of a graviton ($\epsilon^{\mu \nu}_g$ symmetric and traceless), we can neglect  the last three terms in the squared bracket  of Eq.~(\ref{1gra4tacbvv}) and we get 
\ea{
\epsilon^{\mu \nu}_g M_{\mu \nu} (q; k_i )  = \kappa_d   \epsilon^{\mu \nu}_g   \sum_{i=1}^{n} \left[ \frac{k_{i\mu} k_{i\nu}}{k_i q} - i \frac{k_{i\nu} q^\rho J^{(i)}_{\mu \rho} }{2 k_i q} -  i \frac{k_{i\mu} q^\rho J^{(i)}_{\nu \rho} }{2 k_i q}  - \frac{1}{2} \frac{q_{\rho} J_i^{\mu \rho} q_{\sigma} J^{\nu \sigma}_i  }{k_i q} \right]T_n (k_i ) \ , 
\label{Mgravi}
}
which, of course, agrees with the soft theorem for the graviton derived in section 3  of  Ref.~\cite{BDDN}.

In the case of the dilaton, using $\epsilon^{\mu \nu}_d$ given in Eq.~(\ref{epsd}),
one gets instead:
\begin{eqnarray}
 \epsilon^{\mu \nu}_d M_{\mu \nu} (q; k_i)  = \kappa_d
 && \left[  - \sum_{i=1}^{n} \frac{ m_i^2 \left( 1 + q^{\rho}  \frac{\partial}{\partial k_{i}^{\rho}}
+ \frac{1}{2} q^{\rho}   q^{\sigma} \frac{ \partial^2}{ \partial k_{i}^{\rho} \partial k_{ i}^{\sigma}   }  \right)
}{k_i q}     -    \sum_{i=1}^{n} k_{i}^{\rho}   \frac{\partial}{ \partial k_{i}^{\rho }} +2 \right. \nonumber \\
 && \ \ - \left.  \sum_{i=1}^{n}  \left(k_{i\mu} q_{\sigma} \frac{\partial^2}{\partial k_{i\mu} \partial k_{i\sigma}} 
 - \frac{1}{2} (k_i q) \frac{\partial^2}{\partial k_{i\mu} \partial k_{i\mu}} \right) \right]
 T_n (k_i ) \ ,
\label{dilantac}
\end{eqnarray}
where $m_i^2 = -\frac{4}{\alpha'}$ is the squared mass of the closed string tachyon. 
The dilaton contains terms $\Ord( q^{-1} )$ when the other particles are massive, because the three-point amplitude involving a dilaton and two equal particles with mass $m$ is  proportional to $m^2$.

In  the case of the  Kalb-Ramond field we get zero  simply because it is not coupled to $n$ tachyons. This is due to the fact that the closed bosonic string is invariant under the world-sheet parity $\Omega$ that leaves invariant the vertex operators of the tachyon, dilaton and graviton, while changes sign of the vertex operator of $B_{\mu \nu}$. 

The most important result of  our analysis is that   Eq.~\eqref{MMMmunu}, which was derived in Ref.~\cite{BDDN} for  a graviton from Eqs.~\eqref{qqM}, also gives the correct  amplitude for a dilaton when it is saturated with $\epsilon_{d}^{\mu \nu}$ given in Eq.~(\ref{GdB}).  This means that 
extra  gauge invariant terms proportional to $\eta_{\mu \nu}$ do not appear in the matrix $M_{\mu \nu}$ in Eq.~(\ref{1gra4tacbvv}).

We have then studied  the soft behavior of a massless closed string state in an amplitude involving an arbitrary number of other massless closed string states. In this case we have performed the calculation up to the $\Ord (q^0)$, and  for the symmetric part of $M_{\mu \nu}$ we get:
\begin{eqnarray}
M^{\mu \nu}_{S} (q ; k_i , \epsilon_i) = \kappa_d  \sum_{i=1}^{n} \left(\frac{ k_i^{\mu} k_i^{\nu} - \frac{i}{2} k_i^{\nu} q_\rho J_i^{\mu \rho} - \frac{i}{2} k_i^{\mu} q_\rho J_i^{\nu \rho} }{ q k_i}  \right) M_n (k_i , \epsilon_i ) + \Ord (q) \ ,
\label{Msoft}
\end{eqnarray}
where
$ M_n (k_i , \epsilon_i )$ is the amplitude with $n$ massless states,
\begin{eqnarray}
J_i^{\mu \nu} = L_i^{\mu \nu} + S_i^{\mu \nu} + {\bar{S}}^{\mu \nu}_i \ ,
\label{JLS}
\end{eqnarray}
\ea{
L_i^{\mu\nu} =i\left( k_i^\mu\frac{\partial }{\partial k_{i\nu}} -k_i^\nu\frac{\partial }{\partial k_{i\mu}}\right) \, , \ 
S_i^{\mu\nu}=i\left( \epsilon_i^\mu\frac{\partial }{\partial \epsilon_{i\nu}} -\epsilon_i^\nu\frac{\partial }{\partial \epsilon_{i\mu}}\right) \, , \  
{\bar{S}}^{\mu\nu}_i=i\left( {\bar{\epsilon}}_i^\mu\frac{\partial }{\partial {\bar{\epsilon}}_{i\nu}} -{\bar{\epsilon}}_i^\nu\frac{\partial }{\partial {\bar{\epsilon}}_{i\mu}}\right)  \, .
\label{LandS}
}
Note that in the previous expressions  we have written the polarizations of the massless closed string states as a product of the polarizations of two massless open string states, namely  \mbox{$\epsilon^{\mu \nu}_i = \epsilon^{\mu}_i {\bar{\epsilon}}^{\nu}_i$}. 

By saturating Eq.~(\ref{Msoft}) with the polarization of the graviton, one gets the soft graviton behavior, which agrees with what was obtained from gauge invariance in Ref.~\cite{BDDN}. If we instead saturate it with the polarization of the dilaton we get:
\begin{eqnarray}
M_{n+1} = \kappa_d
 \left[ 2 - \sum_{i=1}^{n}  k_{i\mu} \frac{\partial}{\partial k_{i\mu}}  \right] M_n 
+ \Ord (q ) \ ,
\label{softdilathe}
\end{eqnarray}
which agrees with the result obtained in Ref.~\cite{Ademollo:1975pf}.  Note that we do not get any terms of $\Ord(q^{-1})$, as we got for the amplitude with external tachyons, since the external states are now massless.

The soft theorem for the dilaton can be written in a more suggestive way~\cite{Ademollo:1975pf}  by observing that, in general, $M_n$ has the following form:
\begin{eqnarray}
M_n = \frac{8\pi}{\alpha'} \left(\frac{\kappa_d}{2\pi}\right)^{n-2} F_n (\sqrt{\alpha' }k_i ) \, , \
\kappa_d = \frac{1}{2^{\frac{d-10}{4}} } \frac{g_s}{\sqrt{2}} (2\pi)^{\frac{d-3}{2}} (\sqrt{\alpha'})^{\frac{d-2}{2}} \, ,
\label{MnTn}
\end{eqnarray}
where $F_n$ is dimensionless and obviously satisfies the equation:
\begin{eqnarray}
\sum_{i=1}^{n}  k_{i\mu} \frac{\partial}{\partial k_{i\mu}}  F_n = \sqrt{\alpha'} \frac{\partial}{\partial \sqrt{\alpha'}} F_n \,\, .
\label{partTn}
\end{eqnarray}
Using these equations, the soft theorem for the dilaton becomes~\cite{Ademollo:1975pf}:
\begin{eqnarray}
M_{n+1} = \kappa_d \left[ -  \sqrt{\alpha'} \frac{\partial}{\partial \sqrt{\alpha'}} + \frac{d-2}{2} g_s \frac{\partial}{\partial g_s}    \right] M_n + \Ord (q ) \,\, .
\label{softfinale}
\end{eqnarray}
Therefore,  the emission of a dilaton with zero momentum is obtained from the amplitude without a dilaton
by  a simultaneous rescaling of the Regge slope $\alpha'$ and the string coupling constant $g_s$. This is the same rescaling that leaves Newton's constant invariant:
\begin{eqnarray}
\left[ -  \sqrt{\alpha'} \frac{\partial}{\partial \sqrt{\alpha'}} + \frac{d-2}{2} g_s \frac{\partial}{\partial g_s}      \right] \kappa_d =0 \, ,
\label{kappadinva}
\end{eqnarray}
as can be checked from its definition in Eq.~(\ref{MnTn}).

In order to explore the possibility of formulating a soft theorem for the antisymmetric tensor, it is convenient to keep distinct the holomorphic and anti-holomorphic sectors coming from the factorized structure of the vertices in closed string theory.
According to such a separation the amplitude $M_n (k_i , \epsilon_i; {\bar{k}}_i, {\bar{\epsilon}}_i)$, on which the soft operator  acts,  becomes a function of the holomorphic, $k_i$, and anti-holomorphic, ${\bar{k}}_i$, momenta. This separation  is only an intermediary trick of the calculation; the momenta must at the end be identified  as required by the BRST invariance of the theory.  Because of this   splitting, however,  we have   to also introduce the anti-holomorphic angular momentum operator
\begin{eqnarray}
{\bar{L}}_i^{\mu\nu} = i\left( {\bar{k}}_i^\mu\frac{\partial }{\partial {\bar{k}}_{i\nu}} -{\bar{k}}_i^\nu\frac{\partial }{\partial {\bar{k}}_{i\mu}}\right) \, .
\label{Lbarbar}
\end{eqnarray}
In terms of these operators, the soft behavior for the antisymmetric tensor reads:
\ea{
M_{n+1}  &=  -i\epsilon_{q\, \mu\nu}^B \kappa_d \sum_{i=1}^{n}   \left[   \frac{k_i^\nu q_\rho (L_i+S_i)^{\mu\rho}}{q k_i}   -    \frac{k_i^\nu q_\rho (\bar{L}_i+\bar{S}_i)^{\mu\rho}}{q k_i}  \right] {M_n(k_i , \epsilon_i ; {\bar{k}}_i, {\bar{\epsilon}}_i )}\Big |_{k=\bar{k}}  + \Ord (q) \nonumber \\
& = -i\epsilon_{q\, \mu\nu}^B \kappa_d \sum_{i=1}^{n}   \left[    \frac{1}{2} ( L_i - {\bar{L}}_i )^{\mu \nu} + \frac{k_i^{\nu} q_{\rho}}{k_i q} ( S_i - {\bar{S}}_i)^{\mu \rho} \right]{M_n(k_i , \epsilon_i ; {\bar{k}}_i, {\bar{\epsilon}}_i )}\Big |_{k=\bar{k}}  + \Ord (q)   
\, .
\label{Bmunu44}
}
As expected from Weinberg's general argument, we do not get any term of $\Ord (q^{-1})$, corresponding to a long range force, but there are several terms of $\Ord (q^0)$. 
 By construction, Eq. (\ref{Bmunu44}) reproduces the soft behavior of the antisymmetric tensor,  
but it is not a real soft theorem as in the case of the graviton and dilaton because, due to the separation of $k$ and ${\bar{k}}$,  the amplitude $M_n$ is not a physical amplitude.

The paper is organized as follows. In Sect. \ref{tachyon} we discuss the soft behavior for the graviton and the dilaton coupled to $n$ closed string tachyons. In Sect. \ref{massless}  we
turn to the case where all  external states are massless closed string states. Sect. \ref{ConclusionSection} is devoted to some conclusions and outlook.
The details of various calculations are given in the appendix.

\section{One massless  closed string and $n$ tachyons}
\label{tachyon}
\setcounter{equation}{0}

The scattering amplitude involving a massless closed string state and $n$ closed string tachyons is given by~\footnote{In this and in the next section we omit to write the overall normalization factors discussed in the introduction.}:
\begin{eqnarray}
M_{\mu \nu}  \sim \int \frac{\prod_{i=1}^{n} d^2 z_i}{d V_{abc}} \prod_{i<j} |z_i - z_j|^{\alpha' k_i k_j} 
S _{\mu \nu} \, ,
\label{M1n} 
\end{eqnarray}
where
\begin{eqnarray}
S_{\mu \nu}  =  \frac{\alpha'}{2}  \int d^2 z \prod_{\ell=1}^{n} | z- z_{\ell}|^{\alpha' k_{\ell} q}\sum_{i=1}^{n}  \frac{k_{i\mu}}{z- z_i} \sum_{j=1}^{n}  \frac{k_{j\nu}}{{\bar{z}} - {\bar{z}}_j} \, .
\label{Nzbarzqki}
\end{eqnarray}
The quantities $z_i$, with $i=1, \dots, n$, are complex coordinates
parametrizing the insertion on the world-sheet of the vertex operators
associated to  the tachyon states. The coordinate~$z$,Ê without index,Ê is
associated to the massless closed string state.

In order to find a soft operator $\hat{S}$ such that $M_{n+1} = \hat{S} M_n$,
we first need to compute $S_{\mu \nu}$ for small $q$. This calculation is performed in the appendix. For the `diagonal' terms, i.e. terms in the sum where $i=j$, the result is given in Eq.~(\ref{sumIi}) and one gets:
\ea{
&  S^{\rm diag}_{\mu \nu}= 2\pi \sum_{i=1}^{n} k_{i\mu} k_{i \nu} \left[  \alpha' \log \Lambda  + \frac{(\alpha')^2}{2} \sum_{j \neq i} (k_j q) \log^2 |z_i - z_j|  + \frac{1}{k_i q}  \right. \nonumber \\
& \times \left. \left( 1 +\alpha'  \sum_{j \neq i} (k_j q) \log |z_i - z_j| + \frac{(\alpha')^2}{2} 
\sum_{j \neq i} \sum_{k \neq i} (k_j q) (k_k q)  \log|z_i -z_j| \log |z_i - z_k| \right) \right] \, .
\label{sumIix}
}
The integral of the `non-diagonal' terms, i.e. where $i \neq j$ in Eq.~\eqref{Nzbarzqki},  is given by Eqs.~(\ref{ndfinal}) and (\ref{finalnondiag}) 
and is equal to
\begin{eqnarray}
S^{\rm non-diag}_{\mu \nu} = && 2\pi \sum_{i \neq j} \frac{k_{i\mu} k_{j\nu} + k_{i\nu} k_{j\mu}}{2} \left\{  \alpha'  
\left[\log\Lambda-\log|z_i-z_{j}|\right] \right. \nonumber \\
&& +   \frac{(\alpha' )^2}{2}  \left[ \sum_{k \neq i,j} (k_k q)   \left( \log |z_k - z_{i}|  
\log |z_k - z_{j}| \right)  \right. \label{nondigfni} \\
&& \left. \left.- \sum_{k \neq i} (k_k q) \log|z_i - z_{j}| \log |z_k - z_{i}|
- \sum_{k \neq j} (k_k q) \log|z_i - z_{j}| \log |z_k - z_{j}| \right]  \right\} \ .\nonumber 
\end{eqnarray}
This is only the symmetric part of the non-diagonal contribution. The antisymmetric part is zero after integrating over the $n$ complex coordinates $z_i$,  as shown in the appendix. As explained in the introduction, this is a consequence of the $\Omega$ parity invariance, which does not allow the coupling of one Kalb-Ramond field to $n$ tachyons.

Each of the two previous contributions is divergent when $|z|$ goes to infinity, which is why we have introduced a cutoff $\Lambda$.
After summing the two contributions, however, the divergent terms cancel due to momentum conservation  and we are left with a finite expression.  In conclusion, Eq.~(\ref{M1n}) is equal to
\begin{align}
\label{totalexpre}
 M_{\mu \nu} \sim &  2\pi  \int \frac{\prod_{i=1}^{n} d^2 z_i}{d V_{abc}} \prod_{i<j} |z_i - z_j|^{\alpha' k_i k_j}    \\
& \times \left\{  \sum_{i=1}^{n} k_{i\mu} k_{i \nu} \left[    \frac{(\alpha')^2}{2} \sum_{j \neq i} (k_j q) \log^2 |z_i - z_j|  + \frac{1}{k_i q}  \right.  \right. \nonumber \\
& \times \left. \left( 1 +\alpha'  \sum_{j \neq i} (k_j q) \log |z_i - z_j| + \frac{(\alpha')^2}{2} 
\sum_{j \neq i} \sum_{k \neq i} (k_j q) (k_k q)  \log|z_i -z_j| \log |z_i - z_k| \right) \right]  
\nonumber \\ 
& +  \sum_{i \neq j} \frac{k_{i\mu} k_{j\nu} + k_{i\nu} k_{j\mu}}{2} \left[ - \alpha'  \log|z_i-z_{j}|
+   \frac{(\alpha' )^2}{2}  \left( \sum_{k \neq i,j} (k_k q)   \left( \log |z_k - z_{i}|  
\log |z_k - z_{j}| \right)  \right. \right. \nonumber \\
& \left. \left. \left. - \sum_{k \neq i} (k_k q) \log|z_i - z_{j}| \log |z_k - z_{i}|
- \sum_{k \neq j} (k_k q) \log|z_i - z_{j}| \log |z_k - z_{j}| \right) \right]\right\} + \Ord (q^2) \ .\nonumber
\end{align}
It is easy to see that the three terms of order $(\alpha')^0$ and $(\alpha' )^1$ (equivalently $q^{-1}$ and $q^0$) can be written in the following  compact form:
\begin{eqnarray}
2\pi \sum_{i=1}^{n} \left[ \frac{k_{i\mu} k_{i\nu}}{k_i q} - i \frac{k_{i\nu} q^\rho J^{(i)}_{\mu \rho} }{ 2k_i q}  - i \frac{k_{i\mu} q^\rho J^{(i)}_{\nu \rho} }{2k_i q}     \right] \int \frac{\prod_{i=1}^{n} d^2 z_i}{d V_{abc}} \prod_{i \neq j} |z_i - z_j|^{\frac{\alpha'}{2} k_i k_j}  \ ,
\label{01}
\end{eqnarray}
where the last  integral is the amplitude $M_n$ of $n$ closed string tachyons, given with the correct normalization in Eq.~\eqref{Mtachyons},  and
\begin{eqnarray}
J^{(i)}_{\mu \rho} = i \left(  k_{i\mu} \frac{ \partial}{\partial k_{i}^{\rho}} -  k_{i\rho} \frac{\partial}{\partial k_{i}^{\mu}}  \right) \, .
\label{JJJ}
\end{eqnarray}
We are left with the terms of order $(\alpha')^2$  (i.e. $\Ord(q)$)  in Eq.~(\ref{totalexpre}) which read 
\ea{
& 2\pi  \frac{(\alpha')^2}{2}   \int \frac{\prod_{i=1}^{n} d^2 z_i}{d V_{abc}} \prod_{i<j} |z_i - z_j|^{\alpha' k_i k_j}  \nonumber \\
& \times  \left[  \sum_{i=1}^{n} \frac{k_{i\mu} k_{i \nu}}{k_i q} \sum_{j \neq i} \sum_{k \neq i} (k_j q) (k_k q)  \log|z_i -z_j| \log |z_i - z_k|    \right.  \nonumber \\
& + \sum_{i=1}^{n} k_{i\mu} \sum_{j=1}^{n} k_{j\nu} \sum_{k \neq i,j} (k_k q)   \left( \log |z_k - z_{i}|  \log |z_k - z_{j}| \right) \nonumber \\
& \left.  - \sum_{i=1}^{n} k_{i\mu} \sum_{j \neq i} k_{j\nu} \log |z_i - z_j| \left( \sum_{k \neq i} (k_k q)  \log |z_k - z_{i}| + \sum_{k \neq j} (k_k q)  \log |z_k - z_{j}|  \right) \right], 
\label{subsub2}
}
and which can be written as
\begin{eqnarray}
&&\frac{1}{2} \sum_{i=1}^{n} \frac{q^\rho q^\sigma}{k_i q} \left[ k_{i\mu }k_{i\nu} \frac{\partial^2}{\partial k_{i}^{\rho} \partial k_{i}^{\sigma}}  -  k_{i\rho }k_{i\nu} \frac{\partial^2}{\partial k_{i}^{\mu} \partial k_{i}^{\sigma}} - k_{i\mu }k_{i\sigma} \frac{\partial^2}{\partial k_{i}^{\rho} \partial k_{i}^{\nu}} + k_{i\rho }k_{i\sigma} \frac{\partial^2}{\partial k_{i}^{\mu} \partial k_{i}^{\nu}}    \right] \nonumber \\
&& \times 2 \pi      \int \frac{\prod_{i=1}^{n} d^2 z_i}{d V_{abc}} \prod_{i\neq j} |z_i - z_j|^{\frac{\alpha'}{2} k_i k_j}   \ .
\label{expecta}
\end{eqnarray}
It is easy to check that the first term of the previous expression is equal to the first term of Eq.~(\ref{subsub2}), the last term in Eq.~(\ref{expecta}) is equal to the second term of Eq.~(\ref{subsub2}), and finally the third and fourth terms in Eq.~(\ref{expecta}) are equal to the other two terms  of Eq.~(\ref{subsub2}).
Eq.~(\ref{expecta}) can also be written as:
\begin{eqnarray}
&&  \frac{1}{2} \sum_{i=1}^{n} \frac{1}{k_i q} \left[  \Big( (k_i q) ( \eta_{\mu \nu}  q_\sigma - q_\mu \eta_{\nu \sigma} ) -k_{i \mu} q_\nu q_\sigma \Big) \frac{\partial}{\partial k_{i\sigma}} - q^{\rho} 
J_{i \,\mu \rho} q^{\sigma} J_{ i \, \nu \sigma}   \right] 
\nonumber \\
&& 
\times   
  2 \pi   \int \frac{\prod_{i=1}^{n} d^2 z_i}{d V_{abc}} \prod_{i\neq j} |z_i - z_j|^{\frac{\alpha'}{2} k_i k_j}  \ .
\label{subsub3}
\end{eqnarray}
Thus, in total, we get the following soft behavior:
\begin{eqnarray}
M_{\mu \nu} \sim
   &&
   \sum_{i=1}^{n} 
   \left[ \frac{k_{i\mu} k_{i\nu}}{k_i q}  - i \frac{k_{i\nu} q^\rho J^{(i)}_{\mu \rho} }{2k_i q}  - i \frac{k_{i\mu} q^\rho J^{(i)}_{\nu \rho} }{2k_i q} - \frac{1}{2} \frac{q^{\rho} J_{i \,\mu \rho} q^{\sigma} J_{i \,\nu \sigma}  }{k_i q}   + \left(  \frac{1}{2} \left(\eta_{\mu \nu}  q_{\sigma} - q_\mu \eta_{\nu \sigma}\right) \right. \right. \nonumber \\
&& \qquad \left. \left.  - \frac{k_{i \mu} q_\nu q_\sigma  }{ 2k_iq} \right)  \frac{\partial}{\partial k_{i \sigma}} \right] 
2 \pi  \int \frac{\prod_{i=1}^{n} d^2 z_i}{d V_{abc}} \prod_{i \neq j} |z_i - z_j|^{\frac{\alpha'}{2} k_i k_j}   + O(q^2) \ .
\label{1gra4tacbvvv}
\end{eqnarray}  

For the graviton we can forget the last three terms in the squared bracket  and we get:
\begin{eqnarray}
 \epsilon_{g}^{\mu \nu}   M_{\mu \nu} \sim  
 &&
 \epsilon_{g}^{\mu \nu}   \sum_{i=1}^{n} \left[ \frac{k_{i\mu} k_{i\nu}}{k_i q} - i \frac{k_{i\nu} q^\rho J^{(i)}_{\mu \rho} }{k_i q}  - \frac{1}{2} \frac{q^{\rho} J_{i \, \mu \rho} q^{\sigma} J_{i \, \nu \sigma}  }{k_i q} \right]  
 \nonumber \\
&& \times 2\pi \int \frac{\prod_{i=1}^{n} d^2 z_i}{d V_{abc}} \prod_{i \neq j} |z_i - z_j|^{\frac{\alpha'}{2} k_i k_j}   + O(q^2) \ .
\label{gravisoft}
\end{eqnarray}
In the case of the dilaton it is more convenient  to rewrite the squared bracket in Eq.~(\ref{1gra4tacbvvv}) as in the first line of Eq.~(\ref{MMMmunu}). The amplitude for the emission of a soft dilaton is then equal to:
\begin{eqnarray}
  \epsilon^{\mu \nu}_{d}   M_{\mu \nu} \sim  
  &&
  \left[  - \sum_{i=1}^{n} \frac{ m_i^2 \left( 1 + q^{\rho}  \frac{\partial}{\partial k_{i}^{\rho}}
+ \frac{1}{2} q^{\rho}   q^{\sigma} \frac{ \partial^2}{ \partial k_{i}^{\rho} \partial k_{ i}^{\sigma}   }  \right)
}{k_i q}     -    \sum_{i=1}^{n} k_{i}^{\rho}   \frac{\partial}{ \partial k_{i}^{\rho }} +2 \right. \nonumber \\
 && \ \ - \left.  \sum_{i=1}^{n}  \left(k_{i\mu} q_{\sigma} \frac{\partial^2}{\partial k_{i\mu} \partial k_{i\sigma}} 
 + \frac{1}{2} (k_i q) \frac{\partial^2}{\partial k_{i}^{\mu} \partial k_{i\mu}} \right) \right] \nonumber \\
 &&  \times
 2 \pi
  \int \frac{\prod_{i=1}^{n} d^2 z_i}{d V_{abc}} \prod_{i \neq j} |z_i - z_j|^{\frac{\alpha'}{2} k_i k_j} + \Ord(q^2) \, ,  
\label{dilantacc}
\end{eqnarray}
where $m_i^2 = - \frac{4}{\alpha'}$ and we used conservation of the angular momentum $\sum_{i=1}^{n} J_{\mu \nu}^{(i)} =0$.

In conclusion, with an  explicit calculation of the soft dilaton and graviton  behavior in an amplitude with closed string tachyons in the bosonic string, we have shown that both behaviors come from the same amplitude in Eq.~(\ref{1gra4tacbvvv}), which is equal to the one derived in Ref. \cite{BDDN} from gauge invariance and the conditions in Eq.~(\ref{qqM}). This is the first time that   
the universal soft-behavior up to the second subleading order has been obtained from a string amplitude involving an arbitrary number of closed string states.

\section{One soft and $n$ massless closed strings}
\label{massless}
\setcounter{equation}{0}

In this section we consider the amplitude with $n+1$ massless closed string states and we study its behavior in the limit  in which one of the massless states is soft. 

In order to extract the soft behavior, it is convenient to  write the vertex operator  of the massless closed string state in the following form:
\begin{eqnarray}
V(z,\,\bar{z})
&&=\left(i \epsilon \cdot\frac{\partial_z X(z)}{\sqrt{2\alpha'}} e^{i\sqrt{\frac{\alpha'}{2}} k\cdot X(z)} \right) \left(i \bar{\epsilon} \cdot\frac{\partial_{\bar{z}} X(\bar{z})}{\sqrt{2\alpha'}} e^{i\sqrt{\frac{\alpha'}{2}} k\cdot X(\bar{z})} \right) \nonumber \\
&& = \int d\theta  ~e^{i( \theta \epsilon_\mu \partial_z+ \sqrt{\frac{\alpha'}{2}} k_\mu)X^\mu(z)} \int d {\bar{\theta}} ~e^{i( \bar{\theta} \bar{\epsilon}_\nu \partial_{\bar{z}}+ \sqrt{\frac{\alpha'}{2}} k_\nu)X^\nu(\bar{z})} \ ,
\label{introtheta}
\end{eqnarray}
where we assume that  both $\theta$ and $\epsilon$ are Grassmann variables. Then, the amplitude involving $n+1$ massless closed string states is given by
\ea{
 M_{n+1} \sim
 &
 \int \frac{\prod_{i=1}^n d^2z_i\,d^2 z}{dV_{abc}}  \int d \theta  
 \prod_{i=1}^n d\theta_i ~ \langle 0| e^{i( \theta \epsilon^\mu_q \partial_{z}+ \sqrt{\frac{\alpha'}{2}} q^\mu)X_\mu(z)}~\prod_{i=1}^ne^{i( \theta_i \epsilon_i^{\mu_i} \partial_{z_i}+ \sqrt{\frac{\alpha'}{2}} k^{\mu_i}_i)X_{\mu_i}(z_i)}|0 \rangle \nonumber\\
&\times \int d {\bar{\theta}}  
\prod_{i=1}^{n} d {\bar{\theta}}_i \langle 0|  e^{i( \bar{\theta} \bar{\epsilon}^\mu_q \partial_{\bar{z}}+ \sqrt{\frac{\alpha'}{2}} q^\mu)X_\mu(\bar{z})}~\prod_{i=1}^ne^{i( \bar{\theta}_i \bar{\epsilon}_i^{\nu_i} \partial_{\bar{z}_i}+ \sqrt{\frac{\alpha'}{2}} k^{\nu_i}_i)X_{\nu_i}(\bar{z}_i)} |0 \rangle \ .
\label{amplitheta}
}
Here $\epsilon_{q \,  \mu \nu} \equiv \epsilon_{q \, \mu} {\bar{\epsilon}}_{q \, \nu}$ is the polarization of the soft-particle.
Using the contraction:
\begin{eqnarray}
\langle X^\mu (z) X^{\nu} (w) \rangle = - \eta^{\mu \nu} \log (z-w) \ ,
\label{contraction}
\end{eqnarray}
we get
\begin{eqnarray}
M_{n+1} \sim
&&
\int \frac{\prod_{i=1}^n d^2z_i\,d^2 z}{dV_{abc}}  \left\{ \int d \theta 
\prod_{i=1}^n d\theta_i     \langle 0|  \prod_{i=1}^ne^{i( \theta_i \epsilon_i^{\mu_i} \partial_{z_i}+ \sqrt{\frac{\alpha'}{2}} k^{\mu_i}_i)X_{\mu_i}(z_i)} |0 \rangle     \right.     
\nonumber \\
&& \times \left. \exp \left[ \left(\theta \epsilon_q^{\mu} \partial_z + \sqrt{\frac{\alpha'}{2}} q^\mu \right) \sum_{i=1}^{n} \left( \theta_i \epsilon_i^{\mu} \partial_{z_i} + \sqrt{\frac{\alpha'}{2}}  k_i^{\mu} \right) \log (z-z_i) \right] \right\} \nonumber \\
&& \times \left\{ \int d {\bar{\theta}}
 \prod_{i=1}^{n} d {\bar{\theta}}_i  \langle 0|  
\prod_{i=1}^ne^{i( \bar{\theta}_i \bar{\epsilon}_i^{\nu_i} \partial_{\bar{z}_i}+ \sqrt{\frac{\alpha'}{2}} k^{\nu_i}_i)X_{\nu_i}(\bar{z}_i)} |0\rangle \right.
\nonumber \\
&& \times \left. \exp \left[ \left({\bar{\theta}} {\bar{\epsilon}}_q^{\nu} \partial_{\bar{z}} + \sqrt{\frac{\alpha'}{2}} q^\nu \right) \sum_{i=1}^{n} \left( {\bar{\theta}}_i {\bar{\epsilon}}_i^{\nu} \partial_{{\bar{z}}_i} + \sqrt{\frac{\alpha'}{2}}  k_i^{\nu} \right) \log ({\bar{z}}-{\bar{z}}_i) \right] \right\} \, . \qquad \qquad \qquad
\label{amplit6}
\end{eqnarray}
The Grassmann integrals over $\theta$ and ${\bar{\theta}}$ can be performed yielding:
\begin{eqnarray}
M_{n+1} \sim
&& \int \frac{\prod_{i=1}^n d^2z_i }{dV_{abc}}  \left[  \int \prod_{i=1}^n d\theta_i  \,\,
\langle 0|  \prod_{i=1}^ne^{i( \theta_i \epsilon_i^{\mu_i} \partial_{z_i}+ \sqrt{\frac{\alpha'}{2}} k^{\mu_i}_i)X_{\mu_i}(z_i)} |0 \rangle  \right] \nonumber \\
&& \times
\left[ \int \prod_{i=1}^{n} d {\bar{\theta}}_i  \langle 0|  
\prod_{i=1}^ne^{i( \bar{\theta}_i \bar{\epsilon}_i^{\nu_i} \partial_{\bar{z}_i}+ \sqrt{\frac{\alpha'}{2}} k^{\nu_i}_i)X_{\nu_i}(\bar{z}_i)} |0\rangle  \right]   \int d^2 z    \prod_{i=1}^{n} |z- z_i|^{\alpha' q k_i} \nonumber \\
&& \times  \epsilon_q^{\mu} \partial_z \sum_{i=1}^{n} \left( \theta_i \epsilon_i^{\mu} \partial_{z_i} + \sqrt{\frac{\alpha'}{2}}  k_i^{\mu} \right) \log (z-z_i) \exp \left[  \sqrt{\frac{\alpha'}{2}}  \sum_{i=1}^{n} \theta_i (\epsilon_i q) \partial_{z_i} \log (z - z_i)  \right]  \nonumber \\
&& \times {\bar{\epsilon}}_q^{\mu} \partial_{\bar{z}} \sum_{i=1}^{n} \left( {\bar{\theta}}_i {\bar{\epsilon}}_i^{\nu} \partial_{{\bar{z}}_i} + \sqrt{\frac{\alpha'}{2}}  k_i^{\nu} \right) \log ({\bar{z}}-{\bar{z}}_i)  \exp \left[  \sqrt{\frac{\alpha'}{2}}  \sum_{i=1}^{n} {\bar{\theta}}_i ({\bar{\epsilon}}_i q) \partial_{{\bar{z}}_i} \log ({\bar{z}} - {\bar{z}}_i)  \right]  \ .\nonumber \\
\label{ampi8f}
\end{eqnarray}
We can formally write this in two parts:
\begin{eqnarray}
M_{n+1} = M_n * S \ ,
\label{MMS}
\end{eqnarray}
where by $*$ a convolution of integrals is understood, and where
\begin{eqnarray}
S \equiv  
&& 
\int d^2 z \,\, \sum_{i=1}^{n} \left(\theta_i \frac{ (\epsilon_q \epsilon_i)}{(z-z_i)^2} +    \sqrt{\frac{\alpha'}{2}} \frac{(\epsilon_q k_i)}{z-z_i} \right) \sum_{j=1}^{n} \left({\bar{\theta}}_j \frac{ ({\bar{\epsilon}}_q {\bar{\epsilon}}_j)}{({\bar{z}}- {\bar{z}}_j)^2} +    \sqrt{\frac{\alpha'}{2}} \frac{({\bar{\epsilon}}_q k_i)}{{\bar{z}}-{\bar{z}}_i} \right) \nonumber \\
&& \times \exp \left[ - \sqrt{\frac{\alpha'}{2}}  \sum_{i=1}^{n} \theta_i \frac{(\epsilon_i q)  }{z-z_i} \right] \exp \left[ - \sqrt{\frac{\alpha'}{2}}  \sum_{i=1}^{n} {\bar{\theta}}_i \frac{({\bar{\epsilon}}_i q)  }{{\bar{z}}-{\bar{z}}_i} \right]\prod_{i=1}^{n} |z- z_i|^{\alpha' q k_i} 
\label{last3lines}
\end{eqnarray} 
is the part describing the soft particle, with momentum $q$ and
polarizations $\epsilon_q$ and ${\bar{\epsilon}}_q$,
while
\begin{eqnarray}
M_n \sim 
&& \int \frac{\prod_{i=1}^n d^2z_i }{dV_{abc}}  \int \prod_{i=1}^n d\theta_i   \langle 0|  \prod_{i=1}^ne^{i( \theta_i \epsilon_i^{\mu_i} \partial_{z_i}+ \sqrt{\frac{\alpha'}{2}} k^{\mu_i}_i)X_{\mu_i}(z_i)} |0 \rangle
\nonumber \\
&& 
\qquad \qquad \ \ \times 
\int \prod_{i=1}^{n} d {\bar{\theta}}_i   \langle 0|  
\prod_{i=1}^ne^{i( \bar{\theta}_i \bar{\epsilon}_i^{\nu_i} \partial_{\bar{z}_i}+ \sqrt{\frac{\alpha'}{2}} k^{\nu_i}_i)X_{\nu_i}(\bar{z}_i)} |0\rangle 
\nonumber \\
= 
&& \int \frac{\prod_{i=1}^n d^2z_i }{dV_{abc}} \int \left[\prod_{i=1}^n d\theta_i \prod_{i=1}^{n} d {\bar{\theta}}_i \right]   \prod_{i<j} |z_i - z_j |^{\alpha' k_i k_j}  \nonumber \\
&& \times \exp \left[ -\sum_{i<j}  \frac{\theta_i \theta_j}{(z_i - z_j)^2}  (\epsilon_i \epsilon_j) + \sqrt{\frac{\alpha'}{2}} \sum_{i \neq j} \frac{ \theta_i (\epsilon_i k_j) }{z_i - z_j}  \right] \nonumber \\
&& \times \exp \left[- \sum_{i<j}  \frac{{\bar{\theta}}_i {\bar{\theta}}_j}{({\bar{z}}_i - {\bar{z}}_j)^2}  ({\bar{\epsilon}}_i {\bar{\epsilon}}_j) + \sqrt{\frac{\alpha'}{2}} \sum_{i \neq j} \frac{ {\bar{\theta}}_i ({\bar{\epsilon}}_i k_j) }{{\bar{z}}_i - {\bar{z}}_j}  \right] 
\label{nonsoftonly}
\end{eqnarray} 
is the amplitude, without the soft particle, of $n$ massless states with momentum $k_i$ and polarizations $\epsilon_i$ and $\bar{\epsilon}_i$. 

We eventually want to find a soft operator $\hat{S}$ up to order $q^0$  such that \mbox{$\hat{S}M_n = M_n \ast S$}, and thus need to compute $S$ up to the same order. 
We do this by expanding $S$ for  small $q$ and keep terms in the integrand up to the order $q$, since higher orders of the integrand cannot yield terms of order $q^0$ after integration. It is useful then to divide $S$ in four parts:
\begin{eqnarray}
S = \sum_{i=1}^{4} S_i + \Ord(q) \ , 
\label{SSi}
\end{eqnarray}
where the first term defined by
\begin{eqnarray}
&&S_1 = \frac{\alpha'}{2} \int d^2 z \,\, \sum_{i=1}^{n}\frac{(\epsilon_q k_i)}{z-z_i}\sum_{j=1}^{n}  \frac{({\bar{\epsilon}}_q k_j)}{{\bar{z}}-{\bar{z}}_j} \prod_{l=1}^{n} |z- z_l|^{\alpha' q k_l} 
= \epsilon_q^\mu \bar{\epsilon}_q^\nu S_{\mu \nu}
\ ,
\label{I1111}
\end{eqnarray}
is simply given by $S_{\mu \nu}$ in Eq.~(\ref{Nzbarzqki}) already computed for the case of the tachyons in the previous section.

The second term is defined to be the higher-order one of $S_1$ and can
thus be denoted as a convolution (for brevity of notation):
\begin{eqnarray}
S_2 = 
&&S_1 \ast
\left( - \sqrt{\frac{\alpha'}{2}}   \right) 
\sum_{k=1}^{n}  \left( \theta_k  \frac{(\epsilon_k q)  }{z-z_k} + {\bar{\theta}}_k \frac{({\bar{\epsilon}}_k q)  }{{\bar{z}}-{\bar{z}}_k} \right) 
\ ,
\qquad \qquad \qquad \qquad \qquad \quad \
\label{I2222} 
\end{eqnarray} 
Furthermore, the third and the fourth terms are  defined to be
\begin{eqnarray}
S_3 = 
&&\sqrt{\frac{\alpha'}{2}}\int d^2 z \sum_{i=1}^{n} \sum_{j=1}^{n} \left[ \left(  \frac{ \theta_i(\epsilon_q \epsilon_i)}{(z-z_i)^2}\right) \left(\frac{({\bar{\epsilon}}_q k_j)}{{\bar{z}}-{\bar{z}}_j}  \right) +  \left(\frac{(\epsilon_q k_j)}{z-z_j} \right) 
\left( \frac{ {\bar{\theta}}_i({\bar{\epsilon}}_q {\bar{\epsilon}}_i)}{({\bar{z}}- {\bar{z}}_i)^2} \right)
 \right] \nonumber \\
&& \times \prod_{\ell=1}^{n} |z- z_{\ell}|^{\alpha' q k_{\ell}}  \left[ 1-    \sqrt{\frac{\alpha'}{2}}      \sum_{k=1}^{n} \left( \theta_k \frac{(\epsilon_k q)}{z-z_k} +{\bar{\theta}}_k \frac{({\bar{\epsilon}}_k q)  }{{\bar{z}}-{\bar{z}}_k} \right)  \right] 
\ .
\label{I5555}
\\[2mm]
S_4 = 
&&\int d^2 z \sum_{i=1}^{n} \left(\theta_i \frac{ (\epsilon_q \epsilon_i)}{(z-z_i)^2}\right) 
\sum_{j=1}^{n} \left({\bar{\theta}}_j \frac{ ({\bar{\epsilon}}_q {\bar{\epsilon}}_j)}{({\bar{z}}- {\bar{z}}_j)^2}  \right)\prod_{\ell=1}^{n} |z- z_{\ell}|^{\alpha' q k_{\ell}}\nonumber \\
&& \times \left[ 1-    \sqrt{\frac{\alpha'}{2}}      \sum_{k=1}^{n} \left( \theta_k \frac{(\epsilon_k  q)}{z-z_k} +{\bar{\theta}}_k \frac{({\bar{\epsilon}}_k q)  }{{\bar{z}}-{\bar{z}}_k} \right)  \right] \ ,
\label{I4444}
\end{eqnarray}
These terms provide all contributions to the $\Ord(q^0)$.  
The computations to this order are provided in the appendix. 
For $S_1$ we get from Eqs.~(\ref{sumIix}) and (\ref{nondigfni}): 
\begin{eqnarray}
S_1 = 2\pi \epsilon_q^{\mu} {\bar{\epsilon}}_q^{\nu} \left[  \sum_{i=1}^{n} \frac{k_{i\mu} k_{i\nu}}{k_i q } + \alpha'   \sum_{j \neq i} \frac{ k_{i\nu} q^{\rho}}{k_i q} \log |z_i - z_j| \left( k_{i\mu} k_{i\rho} - k_{i\rho} k_{j \mu} \right) \right] + \Ord(q) \, .
\label{S1q0}
\end{eqnarray}
For $S_2$ we get from the appendix Eq.~(\ref{S2final}):
\begin{eqnarray}
S_2 = - 2 \pi \epsilon_q^{\mu} {\bar{\epsilon}}_q^{\nu} \sqrt{\frac{\alpha'}{2}} \sum_{i \neq j} \frac{\theta_i (\epsilon_i q)}{z_i - z_j} \left( \frac{k_{j\mu} k_{i\nu}}{k_i q} -  \frac{k_{j\mu} k_{j\nu}}{k_j q}
\right) + \text{c.c.} + \Ord(q) \ .
\label{S2q0}
\end{eqnarray}
For $S_3$ we get from the appendix Eq.~(\ref{S5final}):
\ea{
 S_3  =&2 \pi \epsilon_{q\mu} {\bar{\epsilon}}_{q \nu} 
  \sum_{i \neq j} 
\left[ 
\sqrt{\frac{\alpha'}{2}}
\frac{(k_j q)\theta_i \epsilon_i^{\mu}}{z_i - z_j}
+ \frac{ (\theta_j \epsilon_j q) (\theta_i \epsilon_i^{\mu})}{(z_i - z_j)^2}
 \right]
  \left ( \frac{k_{i}^{\nu}}{k_i q} - \frac{k_{j}^{\nu}}{k_j q} \right)
  + \text{c.c.}+ \Ord(q) \, .
}
Finally, it turns out that $S_4$ to $\Ord(q^0)$ is zero (cf. Eqs.~\eqref{Iiijj} and \eqref{Iiiii}), i.e.
\ea{
S_4 = 0 + \Ord(q) \ .
}

Collecting all terms, up to the order $\Ord (q^0)$, we get:
\begin{eqnarray}
M_n \ast S  = && M_n \ast 2\pi \epsilon_{q\mu} {\bar{\epsilon}}_{q \nu} \left[  \sum_{i=1}^{n}  \frac{k_{i}^{\mu} k_{i}^{\nu}}{k_i q} + \alpha'   \sum_{j \neq i} \frac{ k_{i}^{\nu} q^{\rho}}{k_i q} \log |z_i - z_j| \left( k_{i}^{\mu} k_{j\rho} - k_{i\rho} k_{j}^{ \mu} \right) \right. \nonumber \\
&&  -  \sqrt{\frac{\alpha'}{2}}   \sum_{j \neq i} 
\left (
\frac{\theta_i (\epsilon_i q)}{z_i - z_j} \left( \frac{k_{j}^{\mu} k_{i}^{\nu}}{k_i q} -  \frac{k_{j}^{\mu} k_{j}^{\nu}}{k_j q} \right)  
 +
 \frac{{\bar{\theta}}_i ({\bar{\epsilon}}_i q)}{{\bar{z}}_i - {\bar{z}}_j} \left( \frac{k_{i}^{\mu} k_{j}^{\nu}}{k_i q} -  \frac{k_{j}^{\mu} k_{j}^{\nu}}{k_j q} \right)  \right )
\nonumber \\
&& + \sqrt{\frac{\alpha'}{2}}   \sum_{j \neq i} \theta_i \epsilon_i^{\mu} k_{i}^{\nu} \frac{k_j q}{k_i q} \frac{1}{z_i - z_j} - \sqrt{\frac{\alpha'}{2}} \sum_{i \neq j} \theta_i \epsilon_{i}^{\mu} k_{j}^{\nu} \frac{1}{z_i - z_j} 
\label{1+2+5} 
\\
&&+ \sqrt{\frac{\alpha'}{2}} \sum_{j \neq i}   {\bar{\theta}}_i {\bar{\epsilon}}_i^{\nu} k_{i}^{\mu} \frac{k_j q}{k_i q} \frac{1}{{\bar{z}}_i - {\bar{z}}_j} - \sqrt{\frac{\alpha'}{2}} \sum_{i \neq j} {\bar{\theta}}_i  {\bar{\epsilon}}_{i}^{\nu} k_{j}^{\mu} \frac{1}{{\bar{z}}_i - {\bar{z}}_j} 
\nonumber \\
&&- \left. \sum_{i \neq j} \frac{ (\theta_j \epsilon_j q) (\theta_i \epsilon_i^{\mu})}{(z_i - z_j)^2} \left( \frac{k_j^{\nu}}{k_j q} - \frac{k_i^{\nu}}{k_i q} \right)   - \sum_{i \neq j} \frac{ ({\bar{\theta}}_j {\bar{\epsilon}}_j q) ({\bar{\theta}}_i {\bar{\epsilon}}_i^{\nu})}{({\bar{z}}_i - {\bar{z}}_j)^2} \left( \frac{k_j^{\mu}}{k_j q} - \frac{k_i^{\mu}}{k_i q} \right)  + \Ord(q) \right] \ .
\nonumber 
\end{eqnarray}
Notice that, if we act with $q_{\mu}$ or  $q_{\nu}$ on  the two-index tensor in the squared bracket of this equation,  we get zero. In other words, this tensor satisfies Eqs. (\ref{qqM}).

In order to obtain also the soft theorem for the antisymmetric tensor, we have to make a step back and slightly modify the  amplitude $M_n$ by introducing, together with the momentum $k$ for the holomorphic part, also a momentum   ${\bar{k}}$ for the anti-holomorphic part~\footnote{Actually this is not strictly needed for the part of the amplitude not containing $\theta_i$ and ${\bar{\theta}}_i$.}:
\begin{eqnarray}
M_n (k_i , \epsilon_i ; {\bar{k}}_i , {\bar{\epsilon}}_i ) \sim  
&& \int \frac{\prod_{i=1}^n d^2z_i }{dV_{abc}} \int \left[\prod_{i=1}^n d\theta_i \prod_{i=1}^{n} d {\bar{\theta}}_i \right]   \prod_{i \neq j} \left[ (z_i - z_j )^{\frac{\alpha'}{4} k_i k_j}  ({\bar{z}}_i - {\bar{z}}_j )^{\frac{\alpha'}{4} {\bar{k}}_i {\bar{k}}_j} \right] \nonumber \\
&& \times \exp \left[ - \frac{1}{2}\sum_{i \neq j}  \frac{\theta_i \theta_j}{(z_i - z_j)^2}  (\epsilon_i \epsilon_j) + \sqrt{\frac{\alpha'}{2}} \sum_{i \neq j} \frac{ \theta_i (\epsilon_i k_j) }{z_i - z_j}  \right] \nonumber \\
&& \times \exp \left[- \frac{1}{2}\sum_{i\neq j}  \frac{{\bar{\theta}}_i {\bar{\theta}}_j}{({\bar{z}}_i - {\bar{z}}_j)^2}  ({\bar{\epsilon}}_i {\bar {\epsilon}}_j) + \sqrt{\frac{\alpha'}{2}} \sum_{i \neq j} \frac{ {\bar{\theta}}_i ({\bar{\epsilon}}_i {\bar{k}}_j) }{{\bar{z}}_i - {\bar{z}}_j}  \right] \, .
\label{nonsoftonlyter7}
\end{eqnarray}
This separation is quite natural in string theory because the string vertex operators are usually written as the factorized product of an holomorphic and anti-holomorphic vertex. Here, we are keeping this separation and the identification of the holomorphic and anti-holomorphic momenta is imposed only at the end of the calculation.  

Let us then also introduce $L$ and ${\bar{L}}$ given by:
\begin{eqnarray}
L_i^{\mu\nu} =i\left( k_i^\mu\frac{\partial }{\partial k_{i\nu}} -k_i^\nu\frac{\partial }{\partial k_{i\mu}}\right)\, , \ 
\bar{L}_i^{\mu\nu} =i\left( {\bar{k}}_i^\mu\frac{\partial }{\partial {\bar{k}}_{i\nu}} -{\bar{k}}_i^\nu\frac{\partial }{\partial {\bar{k}}_{i\mu}}\right)\, ,
\label{LLbar}
\end{eqnarray}
 in analogy with $S$ and ${\bar{S}}$  given by:
\begin{eqnarray}
S_i^{\mu\nu}=i\left( \epsilon_i^\mu\frac{\partial }{\partial \epsilon_{i\nu}} -\epsilon_i^\nu\frac{\partial }{\partial \epsilon_{i\mu}}\right)\, , \ {\bar{S}}_i^{\mu\nu}=i\left( {\bar{\epsilon}}_i^\mu\frac{\partial }{\partial {\bar{\epsilon}}_{i\nu}} -{\bar{\epsilon}}_i^\nu\frac{\partial }{\partial {\bar{\epsilon}}_{i\mu}}\right) \, . 
\label{LandSv}
\end{eqnarray} 

The soft operator that reproduces the  soft behavior in Eq.~(\ref{1+2+5})  is equal to
  \begin{eqnarray}
M_{n+1} = \left( \hat{S}^{(0)} + \hat{S}^{(1)} \right) M_n + \Ord (q )
\, ,
\label{S0+S1}
\end{eqnarray}
where
\begin{eqnarray}
\hat{S}^{(0)} = \epsilon_{ q\, \mu \nu }  \sum_{i=1}^{n}   \frac{ k_{i}^{\mu} k_i^\nu}{qk_i }
\, ,   
\label{S0S1}
\end{eqnarray}
and 
\ea{
S^{(1)} M_n  \sim-i  \epsilon_{q\, \mu \nu }  \sum_{i=1}^{n}\left[    \frac{k_i^\nu q_\rho (L_i+ S_i)^{\mu\rho}}{q k_i} +
 \frac{k_i^\mu q_\rho (\bar{L}_i+\bar{S}_i)^{\nu\rho}}{q k_i}   \right] M_n ( k_i , \epsilon_i ; {\bar{k}}_i , {\bar{\epsilon}}_i )\Big |_{k=\bar{k}}
 \  .
\label{Bmunu43b}
}
The previous notation means that  $k_i$ and   ${\bar{k}}_i$ are kept distinct from each other before acting with $L_i$ and ${\bar{L}}_i$, but are identified at the end of the process.
By symmetrizing and antisymmetrizing one gets:
\ea{
S^{(1)} M_n  \sim
 &-i\epsilon_{q\, \mu\nu}^S \sum_{i=1}^{n} \frac{k_i^\nu q_\rho}{q k_i}\Big[   (L_i+ S_i)^{\mu\rho}+
   (\bar{L}_i+\bar{S}_i)^{\mu\rho}  \Big]M_n ( k_i , \epsilon_i ; {\bar{k}}_i , {\bar{\epsilon}}_i )\Big |_{k=\bar{k}}
   \nonumber\\
 &-i\epsilon_{q\, \mu\nu}^B \sum_{i=1}^{n} \frac{k_i^\nu q_\rho}{q k_i}
\Big[   (L_i+ S_i)^{\mu\rho}-
   (\bar{L}_i+\bar{S}_i)^{\mu\rho}  \Big]
   M_n ( k_i , \epsilon_i ; {\bar{k}}_i , {\bar{\epsilon}}_i )\Big |_{k=\bar{k}} \  ,
\label{Bmunu43a}
}
where
\ea{
\epsilon_{q\mu \nu}^{S} = \frac{ \epsilon_{q\mu} \bar{\epsilon}_{q\nu} +  \epsilon_{q\nu} \bar{\epsilon}_{q\mu}}{2} \ , \ 
\epsilon_{q \mu \nu}^{B} =\frac{ \epsilon_{q\mu} \bar{\epsilon}_{q\nu} -  \epsilon_{q\nu} \bar{\epsilon}_{q\mu}}{2} \ .
}
For the symmetric part one can identify $k_i$ with ${\bar{k}}_i$: $ q_\rho (L_{i}^{\mu\rho}+\bar{L}_i^{\mu\rho})M_n ( k_i , \epsilon_i ; {\bar{k}}_i , {\bar{\epsilon}}_i )|_{k=\bar{k}}= q_\rho L_{i}^{\mu\rho}M_n ( k_i , \epsilon_i ; {{k}}_i , {\bar{\epsilon}}_i )|_{k=\bar{k}} )$ which implies:
\begin{eqnarray}
 S^{(1)} M_n =
 &&-i\epsilon_{q\, \mu\nu}^S    \sum_{i=1}^{n}  \frac{k_i^\nu q_\rho J^{\mu\rho}}{q k_i}  M_n ( k_i , \epsilon_i ; {{k}}_i , {\bar{\epsilon}}_i ) \nonumber\\
   &&-i\epsilon_{q\, \mu\nu}^B \sum_{i=1}^{n} \frac{k_i^\nu q_\rho}{q k_i}
\Big[   (L_i+ S_i)^{\mu\rho}-
   (\bar{L}_i+\bar{S}_i)^{\mu\rho}  \Big]
   M_n ( k_i , \epsilon_i ; {\bar{k}}_i , {\bar{\epsilon}}_i )\Big |_{k=\bar{k}} \ ,
\label{Bmunu44a}
\end{eqnarray}
 where
\begin{eqnarray}
J_i^{\mu\nu}= L_i^{\mu\nu}+S_i^{\mu\nu} + {\bar{S}}_i^{\mu\nu} \, .
\label{JLSb}
\end{eqnarray}
Therefore, if we deal with the graviton or dilaton, we do not need to introduce ${\bar{k}}$, while this is necessary for the antisymmetric tensor.

If we use the polarization of a graviton, given in Eq.~(\ref{epsG}), we get the soft behavior for a graviton in agreement with the result of Ref.~\cite{BDDN}. In the case of the dilaton we get instead:
\begin{eqnarray}
\left( \hat{S}^{(0)} + \hat{S}^{(1)} \right) M_n  
= \left[ 2 - \sum_{i=1}^{n} \left( k_{i\mu} \frac{\partial}{\partial k_{i\mu}} - \frac{q \epsilon_i}{k_i q}  k_{i\mu} \frac{\partial}{\partial \epsilon_{i\mu}} - \frac{q {\bar{\epsilon}}_i}{k_i q}  k_{i\mu}
\frac{\partial}{\partial {\bar{\epsilon}}_{i\mu}} \right) \right] M_n \, ,
\label{dilasoft}
\end{eqnarray}
where we have used conservation of the total angular momentum:
\begin{eqnarray}
\sum_{i=1}^{n} J_{\mu \nu}^{(i)} M_n  =0 \, .
\label{angmomcon}
\end{eqnarray}
The last two terms in Eq.~(\ref{dilasoft}) can be neglected because the amplitude is gauge invariant; i.e. these terms essentially substitute the polarization of a particle with its corresponding momentum, while all other indices are saturated  with their corresponding polarization vectors. Gauge invariance implies that one gets zero. In conclusion, the soft behavior of a dilaton in an amplitude with massless closed string states is given by:
\begin{eqnarray}
M_{n+1}=  \left( \hat{S}^{(0)} + \hat{S}^{(1)} \right) M_n +  \Ord(q) \sim 
 \left[ 2 - \sum_{i=1}^{n}  k_{i\mu} \frac{\partial}{\partial k_{i\mu}}  \right] M_n 
+ \Ord (q ) \, ,
\label{sofdilafin}
\end{eqnarray}
in agreement with the result of Ref.~\cite{Ademollo:1975pf}. 

Finally, in  the case of the Kalb-Ramond field, the term $S^{(0)}$ does not contribute because it is symmetric in the index $\mu$ and $\nu$. One gets only the next term:
\ea{
M_{n+1} &=   \hat{S}^{(1)}  M_n + \Ord(q) 
\nonumber \\
&\sim  
 -i\epsilon_{q\, \mu\nu}^B \sum_{i=1}^{n} \frac{k_i^\nu q_\rho}{q k_i}
\Big[   (L_i+ S_i)^{\mu\rho}-
   (\bar{L}_i+\bar{S}_i)^{\mu\rho}  \Big]
   M_n ( k_i , \epsilon_i ; {\bar{k}}_i , {\bar{\epsilon}}_i )\Big |_{k=\bar{k}} 
+ \Ord (q ) \, .
\label{Bmunu43}
}
 
{Although it is at the present stage not clear how to get the soft operator of the antisymmetric field by directly using its own gauge symmetry, as it has been done for the graviton, it is nevertheless easy to show that it is gauge invariant. Under a gauge transformation for the Kalb-Ramond field,   
$\epsilon_{q \, \mu \nu}^{B}  \rightarrow   \epsilon_{q \, \mu \nu}^{B} + q^{\mu} \chi_{\nu} - q^{\nu} \chi_{\mu}$, the amplitude changes as follows
\begin{eqnarray}
 \hat{S}^{(1)}  M_n  \rightarrow  \hat{S}^{(1)}  M_n  +iq_\rho\chi_\mu \sum_{i=1}^n \Big[(L_i+ S_i)^{\mu\rho}- (\bar{L}_i+\bar{S}_i)^{\mu\rho}\Big]
   M_n ( k_i , \epsilon_i ; {\bar{k}}_i , {\bar{\epsilon}}_i )\Big |_{k=\bar{k}} \, .
\end{eqnarray}
The extra term vanishes as a consequence of the identity
\begin{eqnarray}
\Big. \sum_{i=1}^n (L_i+ S_i)^{\mu\rho}M_n( k_i , \epsilon_i ; {\bar{k}}_i , {\bar{\epsilon}}_i )\Big |_{k=\bar{k}}=\Big. \sum_{i=1}^n (\bar{L}_i+ \bar{S}_i)^{\mu\rho}M_n( k_i , \epsilon_i ; {\bar{k}}_i , {\bar{\epsilon}}_i )\Big |_{k=\bar{k}} \, ,
\end{eqnarray}
which can be proved by a direct calculation, ensuring gauge invariance of the amplitude.
}

\section{Conclusions}
\label{ConclusionSection}
\setcounter{equation}{0}
\vspace{-3mm}

In this paper we have computed the behavior of the  scattering amplitudes of the bosonic string 
involving   massless states; i.e. gravitons, dilatons and Kalb-Ramond antisymmetric tensors, and tachyons when the momentum of one massless particle is very small. In the case of a soft graviton our results  agree, as expected, with the behavior found  in  Ref.~\cite{BDDN} up to terms of $\Ord (q)$ when the other particles are tachyons, and up to terms of $\Ord (q^0 )$ when the other particles are massless.  

We also derived  the soft behavior  for the dilaton and for the two-index antisymmetric tensor, which is not obvious how to obtain  using the  methods developed in Ref.~\cite{BDDN}.  The basic reason why this is instead  possible in string theory, is due to the fact that the amplitudes for the emission of a massless particle are all obtained from the same tensor $M_{\mu \nu}$ by saturating it  with the corresponding polarization vector.  It turns out that the soft behavior  of  $M_{\mu \nu}$ is exactly the one obtained in Ref.~\cite{BDDN}, using just gauge invariance before saturating it with the polarization vector of the graviton.

What one learns from these calculations is that, in the cases examined,  there is a common quantity
$M_{\mu \nu}$, whose  soft  behavior  is determined by imposing the 
conditions in  Eq.~(\ref{qqM}), that provides the soft theorem for all massless states by saturating it with their corresponding polarization vector.  This is also what one gets when applying the rules of BCJ duality~\cite{Bern:2008qj,Bern:2010ue}, according to which one can  obtain  the scattering amplitudes for an extended version of gravity, including the dilaton and the Kalb-Ramond field, from the gluon scattering amplitudes.  
One also knows, however, from Ref.~\cite{Ademollo:1975pf} that this is not the full story for the dilaton, because with open massless states one gets extra terms proportional to $\eta_{\mu \nu}$.

Many things remain to be done. One is to include terms of $\Ord (q^1 )$ in  our analysis 
for the  scattering amplitudes involving only massless states, and to extend it to the case in which massless open strings are involved. Another thing is to extend our analysis to the case where the other particles are arbitrary string states. Finally, one should extend all this to the case of superstring theory and to the loop level.  For the superstring we do not expect drastic changes from the results that we got for the bosonic string and we hope to report on this in a future publication~\cite{DMM2}.

\vspace{-5mm}
\subsection*{Acknowledgments} \vspace{-3mm}
We thank  Zvi Bern, Massimo Bianchi,  Scott Davies, Henrik Johansson,  Josh Nohle, Andrea Guerrieri, Rodolfo Russo and Congkao Wen    for very helpful discussions  and useful comments on the draft. PDV thanks Henrik Johansson for the invitation,  and Uppsala University for the hospitality,  in connection with the informal meeting organized by Henrik  in October 2014 that helped us to clarify the connection of this work with BCJ duality. RM is grateful to the Mainz Institute for 
Theoretical Physics (MITP) for its hospitality  and its partial support during the completion of
this work. MM thanks Andrew Strominger for discussions on the soft behavior of the antisymmetric tensor.

\clearpage
\appendix

\sect{Computational details}
\label{appA}
In this Appendix we lay out the procedure for
computing the integrals in  Eqs.~\eqref{Nzbarzqki} and \eqref{last3lines},
discuss the caveats,
and provide a detailed computation of the quantity in  Eq.~\eqref{Nzbarzqki}
up to  $\Ord (q)$. 
The integrals in  Eq.~\eqref{last3lines} to the $\Ord(q^0)$ can be computed in a similar way (with one exception, which will be discussed), and we thus leave out the details here. We plan to present the computations of $S$ in  Eq.~\eqref{last3lines} up to the $\Ord(q)$ in a future work~\cite{DMM2}.

From Eqs.~\eqref{Nzbarzqki} and \eqref{last3lines} it is evident
that we must compute integrals of the type:
\ea{
I_{i_1 i_2 \ldots}^{j_1 j_2 \ldots} =
\int d^2 z \frac{\prod_{l = 1}^n |z-z_l|^{\alpha' k_l q}}{
(z-z_{i_1})(z-z_{i_2}) \cdots (\bar{z}-\bar{z}_{j_1}) (\bar{z}-\bar{z}_{j_2}) \cdots } \ .
\label{GeneralIntegral}
}
These integrals can generically be evaluated after an expansion in $q$ by a substitution of the form $z \to z_{i} + \rho e^{i\theta}$. 
Note that since we use the convention $d^2 z = 2 d Re(z) d Im (z)$, we
have that $d^2 z = 2 \rho d\rho d\theta$.
It can be useful to substitute further $e^{i\theta} \to \omega$, such that  $\int_{0}^{2\pi}  d \theta \cdots \to \oint d \omega \cdots$,  enabling use of Cauchy's integral formula over the unit circle. Note that the expansion in $q$ of the integrand does not correspond to the same order of the integral: in fact a  term of order $q^n$ of the integrand generically yields a result of the integral of the form $Aq^{n-1} + Bq^n + \cdots$ where $A, B, \ldots$ are the coefficients of integration. Special care must, however, be taken, when dealing with special pole structures such as:
\ea{
\frac{1}{|z-z_i|^2 |z -z_j |^2} \ ,
\label{specialpole}
}
as we shall discuss in a moment.

We note that infrared divergences should not appear, since they will be regulated by $q>0$; i.e. the exponent $\alpha' k_l q$ should not be expanded for those $l$'s for which $z-z_l$ is a simple pole of the integrand. Ultraviolet divergences instead can appear due to partitioning of integrals, but must cancel in the final result.

In terms of Eq.~\eqref{GeneralIntegral} we can write the relevant integrals of this work as follows:
\ea{
S_{1} = &\frac{\alpha'}{2} \epsilon_q^\mu \bar{\epsilon}_q^\nu \sum_{i,j = 1}^n k_{i\mu} k_{j \nu} I_i^j \ ,
\label{AppS1}
\\[2mm]
S_2 =&
-\left ( \frac{\alpha'}{2} \right)^{{3}/{2}} \epsilon_q^\mu \bar{\epsilon}_q^\nu \sum_{i,j,m = 1}^n k_{i\mu} k_{j \nu} \Big (
(\theta_m \epsilon_m q) I_{im}^j + (\bar{\theta}_m \bar{\epsilon}_m q) I_{i}^{jm} \Big ) \ ,
\label{AppS2}
\\[2mm]
S_3 =&
\sqrt{\frac{\alpha'}{2}} 
 \epsilon_{q \, \mu} \bar{\epsilon}_{q \, \nu} \sum_{i,j = 1}^n
\Bigg[ (\theta_i \epsilon_i^\mu) k_j^\nu I_{ii}^j + (\bar{\theta}_i \epsilon_i^\nu) k_j^\mu I_{j}^{ii} 
\nonumber \\
& \qquad \qquad 
- 
\left ( \frac{\alpha'}{2} \right)^{{3}/{2}} 
\sum_{m=1}^n 
\Bigg ( (\theta_i \epsilon_i^\mu) k_j^\nu
\left( (\theta_m\epsilon_m q) I_{iim}^j + (\bar{\theta}_m\bar{\epsilon}_m q) I_{ii}^{jm}\right )
\nonumber \\
& \qquad \qquad \qquad \qquad \qquad \ \ 
+
 (\bar{\theta}_i \bar{\epsilon}_i^\nu) k_j^\mu
\Big( (\theta_m\epsilon_m q) I_{jm}^{ii} + (\bar{\theta}_m\bar{\epsilon}_m q) I_{j}^{iim}\Big ) \Bigg ) \Bigg ] \, .
\label{AppS5}
\\[2mm]
S_4 =& 
 \epsilon_{q \, \mu} \bar{\epsilon}_{q \, \nu} \sum_{i,j = 1}^n
(\theta_i \epsilon_i^\mu) (\bar{\theta}_j \bar{\epsilon}_j^\nu)
\Big (
I_{ii}^{jj} - \sqrt{\frac{\alpha'}{2}} \sum_{m=1}^n 
\left( (\theta_m \epsilon_m q) I_{iim}^{jj} + (\bar{\theta}_m \bar{\epsilon}_m q) I_{ii}^{jjm}
\right ) \Big) \ ,
\label{AppS4}
}
Note that we only need to consider half of the terms, since by complex conjugation:
\ea{
\overline{I_{i_1 i_2 \ldots}^{j_1 j_2 \ldots} } =
I^{i_1 i_2 \ldots}_{j_1 j_2 \ldots}
}
Note also that the upper respectively lower indices of $I_{i_1 i_2 \ldots}^{j_1 j_2 \ldots}$
are totally symmetric. This shows explicitly that $S_2, \ldots, S_4$ are real valued. For $S_1$ it is useful to separate its real part from  its imaginary part. Defining symmetric and antisymmetric polarization tensors as follows:
\ea{
\epsilon_{q}^{S\mu \nu} = \frac{ \epsilon_q^\mu \bar{\epsilon}_q^\nu +  \epsilon_q^\nu \bar{\epsilon}_q^\mu}{2} \ , \ 
\epsilon_{q }^{B\mu \nu} = \frac{ \epsilon_q^\mu \bar{\epsilon}_q^\nu -  \epsilon_q^\nu \bar{\epsilon}_q^\mu}{2} \ ,
}
we can rewrite $S_1$ in the form:
\ea{
S_1 = 
\frac{\alpha'}{2} \epsilon_{q}^{S\mu \nu} \sum_{i, j=1}^n k_{i\mu}k_{j\nu}\frac{I_i^j + \overline{I_i^j}}{2}
+
\frac{\alpha'}{2} \epsilon_{q}^{B\mu \nu} \sum_{i\neq j}^n k_{i\mu}k_{j\nu}\frac{I_i^j -\overline{I_i^j}}{2} \ .
\label{symantisymS1}
}

In the case of Eq.~\eqref{Nzbarzqki}, where $S_1$ is integrated over the punctures $z_i$ of the $n$ tachyons, the integration involves the factor in Eq. (\ref{M1n}) that is  symmetric  in the exchange $z_i \leftrightarrow \bar{z}_i$.
In the above formula of $S_1$, the (anti)symmetric part in $\mu \leftrightarrow \nu$ is also (anti)symmetric in $z_i \leftrightarrow \bar{z}_i$.  Being the antisymmetric part integrated with a symmetric (real) quantity, one gets a vanishing result. 
Therefore in the case of a massless state scattering on $n$ tachyons, only the symmetric part will contribute, i.e.:
\ea{
S_1^{\rm tachyons} = 
\frac{\alpha'}{2} \epsilon_{q}^{S\mu \nu} \sum_{i, j=1}^n k_{i\mu}k_{j\nu}\frac{I_i^j + \overline{I_i^j}}{2} \ .
}
This is, as explained in the introduction, an explicit consequence of the world-sheet parity $\Omega$ invariance that does not allow couplings between an uneven number of Kalb-Ramond fields and tachyons.

The double pole structure of Eq.~\eqref{specialpole}  that appears in $S_4$ requires more care.
In this case  it is convenient to send one pole to infinity by the  projective transformation
\begin{eqnarray}
z\rightarrow z'=\frac{z-z_i}{z-z_j} 
\quad \Longrightarrow \quad
dz=\frac{z_i-z_j}{(z'-1)^2}dz' \ .
\label{cvar}
\end{eqnarray}
Then  one gets
\ea{
I_{ii}^{jj}
= 
\frac{1}{|\zz{i}{j}|^{2-\aqk{i}}}
\int d^2z' \frac{|z'|^{\aqk{i}}}{z'^2} 
\prod_{l\neq i} | (z_i-z_l) - z' (z_j-z_l)|^{\aqk{l}} \ ,
}
where momentum conservation was used to reduce the factor: $|1-z'|^{-\sum_{l=1}^n \aqk{l}} = 1$.
Now, expand in $q$ and consider the first term:
\ea{
\frac{1}{|\zz{i}{j}|^{2}}
\int d^2z' \frac{|z'|^{\aqk{i}}}{z'^2}
=\frac{1}{|\zz{i}{j}|^{2}}
\int_0^\Lambda \frac{d\rho}{\rho^{1-\aqk{i}}}\int_0^{2\pi}d\theta e^{-2i\theta}=0 \ ,
\label{Iiijj}
} 
This shows that  for $i \neq j$ $I_{ii}^{jj}$ in $S_4$ does not contribute to the order $q^0$.

For the diagonal part, one must make use of  the following formula
\ea{
\int d^2 z |z|^{\alpha} |1-z|^{\beta} 
=
\pi \frac{ \Gamma (1+ \frac{\alpha}{2})  \Gamma (1+ \frac{\beta}{2})  \Gamma (-1- \frac{\alpha+\beta}{2})}{ \Gamma (- \frac{\alpha}{2})  \Gamma (- \frac{\beta}{2})
 \Gamma (2+ \frac{\alpha+\beta}{2}) }    \ .
 }
Thus
\ea{
I_{ii}^{ii} &=
\int d^2z |z-z_i|^{\alpha' k_i q - 4} \prod_{j\neq i}^n |z-z_j|^{\alpha' k_j q}
=
 \int d^2w |w|^{\alpha' k_i q - 4} + \Ord(q)
 \nonumber \\
 &=
\pi \frac{\Gamma ( -1 + \frac{\alpha'}{2} k_i q ) \Gamma ( 1)  \Gamma ( 1 - \frac{\alpha'}{2} k_i q)}{ \Gamma (  \frac{\alpha'}{2} k_i q ) \Gamma ( 2 - \frac{\alpha'}{2} k_i q ) \Gamma (0)} + \Ord (q)=0 + \Ord (q) \ ,
\label{Iiiii}
}
demonstrating that   also  $I_{ii}^{ii}$ in $S_4$  does not contribute to the order $q^0$.

\subsection*{Computations for $S_1$ with results for $S_2$ and $S_3$}

{In the last part of this appendix we compute the integral in Eq.~(\ref{Nzbarzqki})  up to the order $\Ord (q)$, also appearing in $S_1$ (Eqs.~\eqref{I1111} and \eqref{AppS1}). The details of the computation also provides the procedure to compute the integrals appearing in $S_2$ and $S_4$, the results of which  to the $\Ord(q^0)$ are quoted in the end.}

We consider first the diagonal part of $I_i^j$, appearing in Eq.~\eqref{AppS1}:
\begin{eqnarray}
I_i^i =
\int d^2 z \prod_{j \neq i} |z-z_j|^{\alpha' k_j q}\, |z-z_i|^{\alpha' k_i q-2}
\ .
\label{d1}
\end{eqnarray}
By writing $z=z_i+\rho\,e^{i\theta}$, one gets
(note that we use the convention $d^2 z = 2 dRe(z) dIm(z)$):
\begin{eqnarray}
I_i^i=2
\int_0^\Lambda d\rho \,\, \rho^{\alpha' k_i q-1}\int_{0}^{2 \pi}
 d\theta  \prod_{j\neq i } |\rho e^{i\theta}+z_i-z_j|^{\alpha' k_j q}
 \ .
\end{eqnarray}
We have introduced a cutoff $\Lambda$ because the integral is divergent for large $\rho$. Expanding the previous expression around  $q=0$ we get
\ea{
I_i^i=
&2
\int_0^\Lambda d\rho \,\, \rho^{\alpha'k_i q-1}\int_{0}^{2 \pi}
 d\theta  \left[ 1 + \sum_{j\neq i} \alpha' (k_j q) \log |z_i - z_j+\rho e^{i\theta} | \right.
 \nonumber \\
 & + \left.  \frac{(\alpha')^2}{2} \sum_{j \neq i} \sum_{k \neq i} (k_j q) (k_k q) \log |z_i - z_j + \rho e^{i\theta} |  \log |z_i - z_k + \rho e^{i\theta} | + \dots  \right] \, .
\label{diago1}
}
It consists of three terms. The first gives:
\ea{
 I_1 &=2 
 \int_0^\Lambda d\rho \,\, \rho^{\alpha'k_i q-1}\int_{0}^{2 \pi}
 d\theta  =  \frac{2\pi}{\alpha' k_i q} \Lambda^{\alpha' k_i q} \nonumber \\
  & = \frac{4\pi}{\alpha'} \left( \frac{1}{k_i q} + \alpha' \log \Lambda  + \frac{(\alpha')^2}{2} (k_i q)  \log^2 \Lambda   \right)
 + \Ord(q^2)\, .
\label{I111}
}
The second term can be written as
\ea{
I_2 &=    
\sum_{j\neq i}   \alpha'  k_j q   
 \int_0^\Lambda d\rho \,\, \rho^{\alpha'k_i q-1} \int_{0}^{2 \pi} d \theta \left( \log (z_i - z_j+\rho e^{i\theta} ) +  \log ( {\bar{z}}_i - {\bar{z}}_j+\rho e^{-i\theta} )    \right) \, .
\label{I222}
}
Dividing the integration regions over $\rho$ allows us to Taylor expand the logarithms yielding:
\begin{eqnarray}
I_2=&& 
\sum_{j\neq i}   \alpha'  k_j q  
\Bigg \{
\int_0^{|z_k-z_i|} d\rho \,\, \rho^{\alpha'k_i q-1}\int_0^{2\pi}d\theta
\nonumber \\
&& \times \left[ \log(z_i-z_j)+\log(1+\frac{ \rho e^{i\theta}}{z_i-z_j}) 
 +\log(\bar{z}_i-\bar{z}_j)+\log(1+\frac{\rho e^{-i\theta}}{\bar{z}_i-\bar{z}_j})\right]\nonumber\\
&&+ \int_{|z_k-z_i|}^\Lambda d\rho\rho^{\alpha' k_i q-1}\int_0^{2\pi}d\theta
\nonumber \\
&&\left[\log(\rho\,e^{i\theta})+\log\left(1+\frac{z_i-z_j}{\rho e^{i\theta}}\right) +
\log(\rho\,e^{-i\theta})+\log\left(1+\frac{\bar{z}_i-\bar{z}_j}{\rho e^{-i\theta}}\right)\right]
\Bigg \}
\nonumber \\
=&&\sum_{j\neq i}   \alpha'  k_j q
\Bigg \{
\int_0^{|z_k-z_i|} d\rho\rho^{\alpha' k_i q-1}\int_0^{2\pi}d\theta \nonumber \\
&& \times \left[
2\log|z_i-z_j|-\sum_{n=1}^\infty\frac{(-1)^n}{n} \frac{\rho^n e^{in\theta}}{(z_i-z_j)^n}
-\sum_{n=1}^\infty\frac{(-1)^n}{n} \frac{\rho^n e^{-in\theta}}{(\bar{z}_i-\bar{z}_j)^n}\right]\nonumber\\
&&+\frac{1}{2} \int_{|z_i-z_j|}^\Lambda d\rho\rho^{\alpha' k_i q-1}\int_0^{2\pi}d\theta \nonumber \\
&& \times \left[2\log\rho
-\sum_{n=1}^\infty\frac{(-1)^n}{n} \frac{(z_i-z_j)^n}{\rho^n e^{in\theta}}-\sum_{n=1}^\infty\frac{(-1)^n}{n} \frac{(\bar{z}_i-\bar{z}_j)^n}{\rho^n e^{-in\theta}}\right]\Bigg\}\, .
\label{lasline}
\end{eqnarray}
Since $ \int_0^{2\pi}d\theta \,\, e^{in\theta} = \delta_{n0}$ the previous expression becomes:
\begin{eqnarray}
I_2 && =  4\pi 
\sum_{j\neq i}   {\alpha'  k_j q} \Bigg\{
 \log |z_i-z_j| \int_0^{|z_i-z_j|} d\rho\,\, \rho^{\alpha' k _i q-1}+
 \int_{|z_i-z_j|}^\Lambda d\rho\,\, \rho^{\alpha' k_i q-1}\log\rho 
\Bigg \}
 \nonumber \\
&& =  
4\pi 
\sum_{j\neq i}  \frac{ k_j q}{ k_i q}  \left[ \Lambda^{\alpha' k_i q}\left(\log\Lambda-\frac{1}{\alpha' k_i q}\right)+ \frac{|z_i - z_j|^{\alpha' k_i q} }{ \alpha' k_i q} \right]  \\
&& = 
4\pi 
\sum_{j\neq i}  \frac{ k_j q}{ k_i q}  \left[  \log |z_i - z_j| + \frac{\alpha'}{2} (k_i q) \log^2 | z_i - z_j| 
+ \frac{\alpha'}{2} (k_i q) \log^2 \Lambda \right] + \Ord(q^2 )\, .
\label{2ling} \nonumber
\end{eqnarray}
Notice that by summing  $I_1$ and $I_2$ the $\log^2 \Lambda$ divergences cancel  due to momentum conservation and to the fact that the soft state is massless
\ea{
I_1 + I_2 \Big |_{\log^2Ê\Lambda} \sim k_i q \log^2Ê\Lambda + \sum_{j\neq i} k_j q \log^2Ê\Lambda = \sum_{j=1}^n k_j q\log^2Ê\Lambda = - q^2 \log^2Ê\Lambda = 0 \ .
}
It remains to compute the last term in Eq.~(\ref{diago1}) given by:
\ea{
I_3 =  (\alpha')^2  
\sum_{j \neq i} \sum_{k \neq i} (k_j q) (k_k q)    
 \int_0^\Lambda d\rho \,\, \rho^{\alpha' k_i q-1}\int_{0}^{2 \pi}
 d\theta  \log |z_i - z_j + \rho e^{i\theta} |  \log |z_i - z_k + \rho e^{i\theta} | \, .
\label{I3xx}
}
In order to evaluate this integral we assume that $|z_i - z_j| > |z_i - z_k|$, and divide the integration range in three regions:
 \begin{eqnarray}
( \Lambda ,   |z_i - z_j|)  \,\,\, ;\,\,\, ( |z_i - z_j|, |z_i - z_k|) \,\,\,;\,\,\, (|z_i - z_k| , 0) 
\label{3regio}
\end{eqnarray}
In the first region we can write the integral part of $I_3$ as follows:
\begin{eqnarray}
&&\frac{1}{4} 
 \int_{|z_i - z_j|}^\Lambda d\rho \,\, \rho^{\alpha' k_i q-1}\int_{0}^{2 \pi} d\theta 
\left[ 2 \log \rho + \log \left(1 + \frac{z_i - z_j}{\rho e^{i\theta}  }\right) +  \log \left(1 + \frac{{\bar{z}}_i - {\bar{z}}_j}{\rho e^{-i\theta}  }\right)  \right]  \nonumber \\
&& \times \left[ 2 \log \rho + \log \left(1 + \frac{z_i - z_k}{\rho e^{i\theta}  }\right) +  \log \left(1 + \frac{{\bar{z}}_i - {\bar{z}}_k}{\rho e^{-i\theta}  }\right)  \right]  \, .
\label{1re}
\end{eqnarray}
After expanding the logarithms the following three terms survive  the integration over $\theta$:
\begin{eqnarray}
&&\frac{\pi}{2}  \int_{|z_i - z_j|}^\Lambda d\rho \,\, \rho^{\alpha' k_i q-1} \left[ 4 \log^2 \rho  +
\sum_{n=1}^{\infty}  \frac{ (z_i - z_j)^n ({\bar{z}}_i - {\bar{z}}_k)^n }{ n^2\rho^{2n}}  
+  \sum_{n=1}^{\infty}  \frac{ (z_i - z_k)^n ({\bar{z}}_i - {\bar{z}}_j)^n }{n^2\rho^{2n}} 
 \right] \, .
 \nonumber \\
\label{3terms}
\end{eqnarray}
Notice that, since $I_3$ already explicitly contains two factors of $q$ and we are only interested in the expression up to order $q$, we should only keep integrands with the power $\rho^{\alpha' k_i q-1}$, because  this is the only way  to get  a $1/q$ contribution. Thus we can readily neglect the last two terms.
Also the first term can be neglected, since the $1/q$ contribution from the lower integration region will cancel the $1/q$ contribution from the upper region after expanding in powers of q. From this we learn the following useful lesson:
\ea{
\int_a^b d \rho \rho^{\alpha' q k_j-n} 
\log^m \rho &= 
\left \{ \begin{array}{ll} 
\Ord(q^0)  &\quad b>a>0  \ \text{and any } n, m 
\\[2mm] 
\frac{1}{\alpha' q k_j} + \Ord(q^0)  &\quad a=0 \ , \ b>0 \ , \ m = 0 \ , \ \text{and } n = 1
\end{array}
\right .
\label{lesson}
}
This means that $I_3$ over the second integration region will also vanish up to order $\Ord(q)$, which we have checked explicitly.

In the third region we can write the integral part of $I_3$ as follows:
\begin{eqnarray}
&&\frac{1}{4}  
\int_{0}^{|z_i -z_k|} d\rho \,\, \rho^{\alpha' k_i q-1}\int_{0}^{2 \pi} d\theta 
\left[  2\log |z_i - z_j|  + \log \left(1 + \frac{ \rho e^{i\theta}  }{z_i - z_j} \right) \right. 
\left. +  
\log \left( 1 + \frac{  \rho e^{-i\theta}  }{{\bar{z}}_i - {\bar{z}}_j} \right)  \right]  \nonumber \\
&&\times \left[  2\log |z_i - z_k|  + \log \left(1 + \frac{ \rho e^{i\theta}  }{z_i - z_k} \right)
 +  
\log \left( 1 + \frac{  \rho e^{-i\theta}  }{{\bar{z}}_i - {\bar{z}}_k} \right)  \right]  \, .
\label{1revvv}
\end{eqnarray} 
Also in this case one gets three terms after integration over $\theta$, but from the above lesson, we see that only the first one contributes to the order $q$:
\ea{
(2\pi) \log|z_i -z_j| \log |z_i - z_k| \int_{0}^{|z_i -z_k|} d\rho \,\, \rho^{\alpha' k_i q-1} =
(2\pi) \frac{\log|z_i -z_j| \log |z_i - z_k|}{\alpha' k_i q} \ .
\label{1ster3}
}
In conclusion $I_3$ reads:
\begin{eqnarray}
I_3 =   \frac{2 \pi \alpha' }{k_i q}  \sum_{j \neq i} \sum_{k \neq i} (k_j q) (k_k q)  \log|z_i -z_j| \log |z_i - z_k| + \Ord(q^2) \, .
\label{I3final}
\end{eqnarray}

In total, the ``diagonal part" of $S_1$
is given by
\ea{
 \frac{\alpha'}{2} I_i^i= 2\pi \Bigg[&\alpha' \log \Lambda  + \frac{(\alpha')^2}{2} \sum_{j \neq i} (k_j q) \log^2 |z_i - z_j|  + \frac{1}{k_i q}  
\left( 1 +\alpha'  \sum_{j \neq i} (k_j q) \log |z_i - z_j|
 \right . \nonumber \\ & \left .  
 + \frac{(\alpha')^2}{2} 
\sum_{j \neq i} \sum_{k \neq i} (k_j q) (k_k q)  \log|z_i -z_j| \log |z_i - z_k| \right) \Bigg]
+ \Ord(q^2) \, .
\label{sumIi}
}
The logarithmic divergence must be cancelled by the remaining non-diagonal terms, which we explicitly demonstrate below.

The non-diagonal terms of $S_1$ are given by the integrals:
\begin{eqnarray}
I_i^{j\neq i} =  \int d^2 z 
\frac{\prod_{m=1}^n |z-z_m|^{\alpha' k_m q}}{(z-z_{i})(\bar{z}-\bar{z}_j)}.
\label{ndiagonal}
\end{eqnarray}
At the lowest order in $q$ of the integrand we get after introducing the variables $z=z_{i} +\rho \,e^{i\theta}$ and $\bar{w}=e^{-i\theta}$:
\begin{eqnarray}
 &&
 \int d^2 z
\frac{|z-z_i|^{\alpha' k_i q}}{(z-z_{i})(\bar{z}-\bar{z}_j)}
 =2\int_0^{\Lambda} \frac{d\rho}{\rho^{1-\alpha' k_i q}} \oint \frac{id\bar{w}}{\bar{w}+\frac{\bar{z}_{i} -\bar{z}_j}{\rho}}
 \nonumber \\
&&=4 \pi   \int_{|z_i -z_j|}^{\Lambda} \frac{d\rho}{\rho^{1-\alpha' k_i q}} 
= 4\pi \left ( \frac{\Lambda^{\alpha' k_i q}- |z_i-z_j|^{\alpha' k_i q}}{\alpha' k_i q} \right )
\label{ndfinal}
\\
&&
=
4\pi \left(\log\Lambda-\log|z_i-z_{j}|
+ \frac{\alpha k_i q}{2} \log^2 \Lambda -
\frac{\alpha k_i q}{2} \log^2 |z_i-z_j|
\right) + \Ord(q^2) \, ,
\nonumber
\end{eqnarray}

In order to evaluate the next term in the expansion we assume that $|z_i-z_m| \leq |z_i-z_j|$ and denote
\begin{eqnarray}
\sum_{m\neq i} \alpha'q k_m\int d^2 z\frac{ |z-z_i|^{\alpha'q k_i}\log|z-z_m| }{(z-z_i)(\bar{z}-\bar{z}_j)}\equiv \sum_{m\neq i} \alpha'q k_m {\cal I}^1_{ij} \, .
\end{eqnarray}
Then using the same substitution as before, we get
\begin{eqnarray}
{\cal I}^1_{ij}
&&= i \int_0^{|z_i-z_m|} d\rho\rho^{\alpha'q k_i-1}\oint d\bar{w}\frac{\log|z_i-z_m|^2+\log\left( 1+\frac{\rho}{\bar{w}(z_i-z_m)}\right)+\log\left(1+\frac{\rho \bar{w}}{\bar{z}_i-\bar{z}_m}\right)}{\bar{w} +\frac{\bar{z}_i-\bar{z}_j}{\rho}}\nonumber\\
&&+i \int_{|z_i-z_m|}^\Lambda  d\rho\rho^{\alpha'q k_i-1}\oint d\bar{w}\frac{
2\log\rho +\log\left(1+\frac{\bar{w}(z_i-z_m)}{\rho}\right)+\log\left(1+\frac{\bar{z}_i-\bar{z}_m}{\rho \bar{w}}\right)}{\bar{w} +\frac{\bar{z}_i-\bar{z}_j}{\rho}}
\ .
\label{s1}
\end{eqnarray} 
By expanding the logarithms we see
that only the second term in the first integral has poles on the contour and is therefore nonvanishing. In the second integral we have to separate the region $[|z_i-z_m|,\,|z_i-z_j|]$ from $[|z_i-z_j|,\,\Lambda ]$. In the first region only the last term has a nonvanishing residue. Thus:
\begin{eqnarray}
&&{\cal I}^1_{ij}=-i  \int_0^{|z_i-z_m|} d\rho\rho^{\alpha'q k_i-1}\oint d\bar{w}\sum_{n=1}^\infty \frac{(-1)^n}{n} \frac{\rho^n}{(z_i-z_m)^n} \left[\frac{1}{\bar{w}^n\left( \bar{w} +\frac{\bar{z}_i-\bar{z}_j}{\rho}\right)}\right]\nonumber\\
&&-i \int_{|z_i-z_m|}^{|z_i-z_j|} d\rho\rho^{\alpha'q k_i-1}\oint d\bar{w}\sum_{n=1}^\infty \frac{(-1)^n}{n} \frac{(\bar{z}_i-\bar{z}_m)^n}{\rho^n}\left[\frac{1}{\bar{w}^n\left( \bar{w} +\frac{\bar{z}_i-\bar{z}_j}{\rho}\right)}\right]
\\
&&+i \int_{|z_i-z_j|}^\Lambda  d\rho\rho^{\alpha'q k_i-1}\oint d\bar{w}\frac{
2\log\rho -\sum_{n=1}^\infty \frac{(-1)^n}{n}\frac{\bar{w}^n(z_i-z_m)^n}{\rho^n}-\sum_{n=1}^\infty \frac{(-1)^n}{n}\frac{(\bar{z}_i-\bar{z}_m)^n}{\rho^n \bar{w}^n}}{\bar{w} +\frac{\bar{z}_i-\bar{z}_j}{\rho}}
\ .
\nonumber 
\end{eqnarray}
The residue formulas of the nonsimple poles read:
\ea{
\text{Res}_{\bar{w}=0}\left[\frac{1}{\bar{w}^n\left( \bar{w} +\frac{\bar{z}_i-\bar{z}_j}{\rho}\right)} \right ]&=
-\left (-\frac{\rho}{\bar{z}_i-\bar{z}_j} \right )^n \ ,
\\
\text{Res}_{\bar{w}=-\frac{\bar{z}_i-\bar{z}_j}{\rho}}\left[\frac{1}{\bar{w}^n\left( \bar{w} +\frac{\bar{z}_i-\bar{z}_j}{\rho}\right)} \right ]&=
\left (-\frac{\rho}{\bar{z}_i-\bar{z}_j} \right )^n \ ,
}
showing that the contour integral over the last term of the last integral vanishes.
It follows by integration and then expansion in $q$ that:
\begin{eqnarray}
{\cal I}^1_{ij}
=&&\pi \sum_{n=1}^\infty \frac{1}{n^2}\frac{(\bar{z}_i-\bar{z}_m)^n}{(\bar{z}_i-\bar{z}_j)^n} 
+2\pi \sum_{n=1}^\infty \frac{1}{n}\frac{(\bar{z}_i-\bar{z}_m)^n}{(\bar{z}_i-\bar{z}_j)^n}
\log\frac{ |z_i-z_j|}{|z_i-z_m|}\nonumber\\
&&+ 2\pi\left[ \log^2\Lambda -\log^2|z_i-z_j| \right]
-\pi \sum_{n=1}^\infty\frac{1}{n^2}\frac{(z_i-z_m)^n}{(z_i-z_j)^{n} }
+O(q)
\nonumber \\
=&&
2\pi \log^2\Lambda -2\pi\log^2|z_i-z_j| 
-2\pi \log\frac{\bar{z}_m-\bar{z}_j}{\bar{z}_i-\bar{z}_j}\log\frac{ |z_i-z_j|}{|z_i-z_m|}
\nonumber\\
&&
+\pi{\rm Li}_2\left( \frac{\bar{z}_i-\bar{z}_m}{\bar{z}_i-\bar{z}_j}\right)
-\pi {\rm Li}_2\left(\frac{z_i-z_m}{z_i-z_j }\right)+O(q) \, ,
\end{eqnarray}
where the Dilogarithmic function was introduced:
\begin{eqnarray}
{\rm Li}_2(z)= \sum_{k=1}^\infty \frac{z^k}{k^2} \ .
\end{eqnarray}
From momentum conservation and $q^2=0$ it follows that the first two terms yield
\begin{eqnarray}
2\pi \sum_{m\neq i}\alpha' q k_m \left ( \log^2\Lambda - \log^2|z_i-z_j| \right)
=-2\pi \alpha' q k_i \left (\log^2\Lambda- \log^2|z_i-z_j| \right ) \, ,
\end{eqnarray}
 which cancel the last two terms in Eq.~\eqref{ndfinal}.
 Thus, summing up, we have found:
\begin{eqnarray}
I_{i}^{j\neq i}=&&4\pi \log\Lambda -4\pi \log|z_i-z_j|
-2\pi \sum_{m\neq i,j}\alpha' q k_m\log\frac{\bar{z}_m-\bar{z}_j}{\bar{z}_i-\bar{z}_j}\log\frac{ |z_i-z_j|}{|z_i-z_m|}
\nonumber\\
&&
+\pi\sum_{m\neq i,j}\alpha' q k_m
\left [
{\rm Li}_2\left( \frac{\bar{z}_i-\bar{z}_m}{\bar{z}_i-\bar{z}_j}\right)-{\rm Li}_2\left(\frac{z_i-z_m}{z_i-z_j }\right)
\right ]
+O(q^2) \, .
\label{finalnondiag}
\end{eqnarray}
Notice that the last line involving the Dilogarithms is purely imaginary and will thus only contribute to the antisymmetric part of $S_1$ according to Eq.~\eqref{symantisymS1}.
The third term can be rewritten as a sum of real and imaginary parts, such that the real part of our result is:
\ea{
&{\cal R}(I_i^{j\neq i} )= 4\pi \log\Lambda -4\pi \log|z_i-z_j|+2\pi\sum_{m\neq i,j}\alpha' q k_m\log|{z}_m-{z}_j|\log|z_i-z_m|\\
&-2\pi \sum_{m\neq j}\alpha' q k_m\log|{z}_m-{z}_j|\log|z_i-z_j|-2\pi\sum_{m\neq i}\alpha' q k_m\log|{z}_i-{z}_j|\log|z_i-z_m| \ ,
\nonumber
}
where momentum conservation is  used to rewrite $\sum_{m\neq i,j} q\cdot k_m = -q \cdot k_i - q\cdot k_j$. The imaginary part, instead, is  equal to:
\ea{
{\cal I}(I_i^{j\neq i} )=\pi\sum_{m\neq i,j}\alpha' q k_m
\Bigg [&{\rm Li}_2\left( \frac{\bar{z}_i-\bar{z}_m}{\bar{z}_i-\bar{z}_j}\right)-{\rm Li}_2\left(\frac{z_i-z_m}{z_i-z_j }\right)
\nonumber\\
& 
+ 
\log\frac{|z_i-z_j|}{|z_i-z_m|}\log\left(\frac{z_m-z_j}{\bar{z}_m-\bar{z}_j}
\frac{\bar{z}_i-\bar{z}_j}{z_i-z_j}\right) \Bigg ] \, .
}
In conclusion, we have found that $S_1$  in Eq.~\eqref{symantisymS1}, up to the $\Ord ( q^1 )$,  is equal to
\ea{
S_1=&
2\pi \epsilon_{q}^{S\mu \nu} \sum_{i=1}^n k_{i\mu}k_{i\nu}
 \Bigg[\alpha' \log \Lambda  + \frac{(\alpha')^2}{2} \sum_{j \neq i} (k_j q) \log^2 |z_i - z_j|  
 \nonumber \\
&
 + \frac{1}{k_i q}  
\left( 1 +\alpha'  \sum_{j \neq i} (k_j q) \log |z_i - z_j|  
 + \frac{(\alpha')^2}{2} 
\sum_{j \neq i} \sum_{k \neq i} (k_j q) (k_k q)  \log|z_i -z_j| \log |z_i - z_k| \right) \Bigg]
\nonumber \\
&
+
2\pi \epsilon_{q}^{S \mu \nu} \sum_{i \neq j}^n k_{i\mu}k_{j\nu}
\alpha' \Bigg[ \log\Lambda -\log|z_i-z_j|+\frac{\alpha' }{2}\sum_{m\neq i,j}(q k_m)\log|{z}_m-{z}_j|\log|z_i-z_m| \nonumber\\
&- \frac{\alpha' }{2}\sum_{m\neq j} (q k_m)\log|{z}_m-{z}_j|\log|z_i-z_j|-\frac{\alpha' }{2}\sum_{m\neq i}(q k_m)\log|{z}_i-{z}_j|\log|z_i-z_m| \Bigg]
\nonumber \\
&
+
2\pi \epsilon_{q}^{B\mu\nu} \sum_{i\neq j\neq m}^n k_{i\mu}k_{j\nu}
\left (\frac{\alpha'}{2}\right )^2  (q k_m) 
\Bigg [{\rm Li}_2\left( \frac{\bar{z}_i-\bar{z}_m}{\bar{z}_i-\bar{z}_j}\right)-{\rm Li}_2\left(\frac{z_i-z_m}{z_i-z_j }\right)
\nonumber\\
& 
+
\log\frac{|z_i-z_j|}{|z_i-z_m|}\log\left(\frac{z_m-z_j}{\bar{z}_m-\bar{z}_j}
\frac{\bar{z}_i-\bar{z}_j}{z_i-z_j}\right) \Bigg ]
+ \Ord(q^2)
\ .
}
It is evident that the logarithmic divergences cancel after using momentum conservation twice on the non-diagonal terms: 
\ea{
\sum_{i\neq j}^n k_{i\mu}k_{j\nu} = 
-\sum_{i =1}^n k_{i\mu}k_{i \nu} + q_\mu q_\nu 
 \ . 
}
The term $ q_\mu q_\nu$ can be neglected because it vanishes in any case when contracted with $\epsilon_{\mu \nu}$.

 Thus, we have derived   
Eq.~\eqref{totalexpre} 
for the scattering with tachyons, 
where the antisymmetric part vanishes 
after the integration over the punctures $z_i$ of the tachyons,
and 
Eq.~\eqref{S1q0}, evaluated only to order $q^0$.

Finally, $S_2$ and $S_3$ are obtained to the order $q^0$ by repeating  exactly the same  procedure  followed  to compute   $S_1$. In so doing we find the following results:
\ea{
S_2 =&  - 2\pi  \epsilon_{q \mu } {\bar{\epsilon}}_{q \nu}\sqrt{\frac{\alpha'}{2}} 
 \sum_{i \neq j}
 \left[   \frac{   \theta_i  (\epsilon_i q)   k_j^{\mu}   }{z_i - z_j} 
\left ( \frac{  k_i^{\nu} }{ k_i q} - \frac{ k_j^{\nu} }{ k_j q } \right )
+   \frac{  {\bar{ \theta}}_i  ({\bar{\epsilon}}_i q)   k_j^{\nu}}{ 
{\bar{z}}_i - {\bar{z}}_j } 
\left ( \frac{  k_i^{\mu}}{ k_i q} - \frac{ k_j^{\mu} }{ k_j q } \right ) \right] + \Ord ( q) \, ,
\label{S2final}
\\
S_3  =&  2\pi \epsilon_{q\,\mu} \bar{\epsilon}_{q\,\nu}
 \sum_{i \neq j} 
\left[ 
\sqrt{\frac{\alpha'}{2}}
\frac{(k_j q)\theta_i \epsilon_i^{\mu}}{z_i - z_j} \left (
 \frac{k_{i}^{\nu}}{k_i q} - \frac{k_{j}^{\nu}}{k_j q} \right)
+  \sqrt{\frac{\alpha'}{2}}\frac{(k_j q){\bar{\theta}}_i {\bar{\epsilon}}_i^{\nu} }{z_i - z_j} \left (
 \frac{k_{i}^{\mu}}{k_i q} - \frac{k_{j}^{\mu}}{k_j q} \right)\right. \nonumber \\
& \left. 
+ \frac{ (\theta_j \epsilon_j q) (\theta_i \epsilon_i^{\mu})}{(z_i - z_j)^2} \left (
 \frac{k_{i}^{\nu}}{k_i q} - \frac{k_{j}^{\nu}}{k_j q} \right) 
 +  \frac{ ({\bar{\theta}}_j {\bar{\epsilon}}_j q) ({\bar{\theta}}_i {\bar{\epsilon}}_i^{\nu})}{({\bar{z}}_i - {\bar{z}}_j)^2} \left (
 \frac{k_{i}^{\mu}}{k_i q} - \frac{k_{j}^{\mu}}{k_j q} \right)  \right] + O (q) Ê\, .
\label{S5final}
}
We have already shown at the beginning of the appendix that $S_4$ vanishes to the order $q^0$. This ends the computations of this paper. We plan to compute and discuss $S$ to the order $q$ in a future work\cite{DMM2}.

\end{document}